\documentclass[12pt]{JHEP3}
\usepackage{pb-diagram,lamsarrow,pb-lams,amsfonts,amssymb,amsthm}

\usepackage{amsmath,epsfig}
\def\be{\begin{equation}}
\def\ee{\end{equation}}
\def\ba{\begin{eqnarray}}
\def\ea{\end{eqnarray}}

\def\be{\begin{equation}}
\def\ee{\end{equation}}
\def\bea{\begin{eqnarray}}
\def\eea{\end{eqnarray}}

\def\yzero{\smash{\hbox{$y\kern-4pt\raise1pt\hbox{${}^\circ$}$}}}

\def\m{\mu}

\def\beq{\begin{equation}}
\def\eeq{\end{equation}}
\def\beqa{\begin{eqnarray}}
\def\eeqa{\end{eqnarray}}

\def\-{\hphantom{-}}

\def\s2{\frac{1}{\sqrt2}}

\def\beq{\begin{equation}}
\def\eeq{\end{equation}}
\def\beqa{\begin{eqnarray}}
\def\eeqa{\end{eqnarray}}

\def\IF{\relax{\rm I\kern-.18em F}}
\def\II{\relax{\rm I\kern-.18em I}}
\def\IP{\relax{\rm I\kern-.18em P}}
\def\IC{\relax\hbox{\kern.25em$\inbar\kern-.3em{\rm C}$}}
\def\IR{\relax{\rm I\kern-.18em R}}

\def\Dsl{\,\raise.15ex\hbox{/}\mkern-13.5mu D} 
\def\IZ{Z\kern-.4em  Z}

\title{Supermembrane origin of type II gauged supergravities in 9D}

\author{M. P. Garc\'\i a del Moral $^1$, J. M. Pe\~na $^2$, A. Restuccia $^3$\,\footnote{E-mail:
\emph{garciamormaria@uniovi.es;jpena@ciens.fisica.ucv.ve;arestu@usb.ve}}\\
$^1$ Departamento de F\'\i sica, Universidad de Oviedo, Avda Calvo
Sotelo S/n. Oviedo, Espa\~na
\\
$^2$ Departamento de F\'\i sica, Facultad de Ciencias,\\
 Universidad Central de Venezuela,
 A.P. 47270, Caracas 1041-A, Venezuela\\
$^3$ Departamento de F\'\i sica, Universidad de Antofagasta, Aptdo 02800, Chile \\
 $\&$ Departamento de F\'\i sica, Universidad Sim\'on Bol\'\i var\\
Apartado 89000, Caracas 1080-A, Venezuela}

\abstract{The M-theory origin of the IIB gauged supergravities in nine dimensions, classified according to the inequivalent classes of monodromy, is shown to exactly corresponds to the  global description of the supermembrane with central charges. The global description is a realization of the sculpting mechanism of gauging (arXiv:1107.3255) and it is associated to particular deformation of fibrations. The supermembrane with central charges may be formulated in terms of sections on symplectic torus bundles with $SL(2,\mathbb{Z})$ monodromy. This global formulation corresponds to the gauging of the abelian subgroups of $SL(2,Z)$ associated to monodromies acting on the target torus. We show the existence of the trombone symmetry in the supermembrane formulated as a non-linear realization of the $SL(2,\mathbb{Z})$ symmetry and construct its gauging in terms of the supermembrane formulated on an inequivalent class of symplectic torus fibration.  The supermembrane also exhibits invariance under T-duality and we find the explicit T-duality transformation. It has a natural interpretation in terms of the cohomology of the base manifold and the homology of the target torus. We conjecture that this construction also holds for the IIA origin of gauged supergravities in 9D such that the supermembrane becomes the origin of all type II supergravities in 9D. The geometric structure of the symplectic torus bundle goes beyond the classification on conjugated classes of $SL(2,\mathbb{Z})$. It depends on the elements of the coinvariant group associated to the monodromy group. The possible values of the (p,q) charges on a given symplectic torus bundle are restricted to the corresponding equivalence class defining the element of the coinvariant group.}

\preprint{FPAUO-12/02}

\begin{document}
\section{Introduction}
The M-theory origin of gauged supergravities is a interesting open
problem. The aim of this paper is to show that the 11D supermembrane
compactified on a torus is the M-theory origin of \underline{\bf
all} supergravities in 9D: not only the maximal supergravity
\cite{julia} but also the gauged sector [\cite{tp}-\cite{etjjto}].  In the
picture we propose, there are two well-differentiated sectors: The
first one is associated to trivial compactifications of the
supermembrane on a 2-torus, its low energy limit corresponds to the
$N=2$ maximal supergravity in 9D, and globally it corresponds to a
trivial symplectic torus bundle. The second sector corresponds to a formulation on a nontrivial symplectic torus bundle. It may occur because a nontrivial monodromy or even in the case of trivial monodromy (the identity) because of a nontrivial cohomology class of the base manifold. The central charge condition is exactly the condition of non-trivial cohomology. The supermembrane with nontrivial central charges corresponds to this sector (\cite{mrt, mor}). In particular we will analyze the formulation on a symplectic torus bundle with nontrivial monodromy. From the physical point of view, the consequence of being a nontrivial cohomology, is very relevant. The spectrum of the hamiltonian becomes discrete with finite multiplicity. By this we refer to the spectrum of the exact hamiltonian, not only to its semiclassical approximation.


It is well-established that the 11D supergravity equations of motion
appear as a consequence of imposing kappa symmetry to the supermembrane action formulated on a general background. This supports the conjecture that the low energy
description of the supermembrane is the 11D
supergravity\footnote{Indeed this conjecture means that the groundstate
of the 11D supermembrane corresponds to the supermultiplet
associated to the 11D supergravity, though, a rigorous proof of this
difficult open problem is still lacking. }. The maximal dimension for
gauged supergravities is 9D. There are four different classes of
gauging appearing in type IIB gauged supergravities in 9D as was
initially established  by \cite{tp},\cite{patrick}. If we include
also the deformations coming from the type IIA sector, there are four
more, but only seven of them are independent deformations and they constitute the type II 9D gauged
supergravity \cite{bergshoeff}, where it is also included the
gauging of scaling symmetries
\cite{samtleben-trombone},\cite{riccioni}. Very recently the most
general gaugings in 9D (expressed in the tensor embedding formalism \cite{dwsn}, \cite{dwst}), have been found in \cite{etjjto}.

Nowadays, the double field theory has become a interesting arena to
try to realize in a bottom-up approach, some of  the properties of
string theory. It is a global approach that describe sigma models
with double coordinates on a $T^{2d}$ torus fibrations such that the
transition functions will be evaluated in the T-duality group
$O(d,d,\mathbb{Z})$. The type II realization has been done recently
in \cite{osb,os}. The proposed action is such that it is invariant
under duality transformations. In 9D the duality transformations
correspond to $SL(2,\mathbb{Z})\times Z_2$ \cite{hull-townsend}.

There is evidence that string theory can be consistently defined in
non-geometric backgrounds in which the transition functions between
coordinate patches involve not only diffeomorphisms and gauge
transformations but also duality transformations
\cite{4hull04},\cite{3hull99}. Some global aspects of T-duality in String theory were formerly analyzed in \cite{lozano}, and more recently  by
\cite{24hull-tfolds}. Such backgrounds can arise from
compactifications with duality twists \cite{dh2} or from
acting on geometric backgrounds with fluxes with T-duality
\cite{3hull99}, \cite{4hull04}, \cite{stw}. In special cases, the
compactifications with duality twists are equivalent to asymmetric
orbifolds which can give consistent string backgrounds
\cite{12flournoyw}, \cite{13kawai},\cite{18cederwall}, \cite{dhgt}. In this
type of compactifications, T-folds are constructed by using strings
formulated on a doubled torus $T^{2n}$ with n-coordinates conjugate
to the momenta and the other n-coordinates conjugate to the winding
modes \cite{4hull04}, plus a constraint to guarantee the correct
number of propagating degrees of freedom.

T-duality transformation at the worldsheet level were studied in \cite{cvetic-stelle}.
The relation of duality and M-theory was also analyzed in \cite{abou}.  In \cite{3hull99,4hull04} it was argued that a fundamental
formulation of string/M-theory should exist in which the T- and
U-duality symmetries are manifest from the start.  In particular, it
was argued that many massive, gauged supergravities cannot be
naturally embedded in string theory without such a framework
\cite{stw}, \cite{6hullreid},\cite{reids},
 \cite{7samtleben}. Examples of
generalized T-folds can be obtained by constructing torus fibrations
over base manifolds with non-contractible cycles. However, in spite of these important advances, up to
our knowledge, a full-fledged realization of these ideas in terms of
worldvolume theories in M-theory is still lacking.

The aim of this paper is to prove that the action of the
Supermembrane  with nontrivial central charges, whose local
structure was given in \cite{mrt},\cite{mor}, may be
globally defined in terms of sections of a symplectic torus bundle
with nontrivial monodromy characterizing at low energies the
gaugings of the type II supergravities. This global description was derived following the sculpting mechanism in \cite{sculpting}. Earlier attempts to establish the connection between the gauging of the supermembrane and that of 9D gauged supergravities can be found in \cite{nisraj},\cite{nisraj2}. \newline
We prove it in the context
of IIB monodromies. The supermembrane formulation on a symplectic torus bundle with monodromy has all the geometrical structure required to derive at low energies the IIB gauged supergravities in 9D. At the level of the supermembrane we are gauging the abelian subgroups of $SL(2,\mathbb{Z})$,  the group of isotopy classes of symplectomorphisms or equivalently area preserving diffeomorphims. It is then natural to think that type IIB gauge supergravities can only interact with the corresponding class of gauged supermembranes in this work. According to
the  inequivalent classes of monodromies, more precisely, to the elements of the coinvariant group of the given monodromy, there is a classification
of the corresponding symplectic torus bundles that describe globally
the supermembrane. The monodromy is given as a representation of the
fundamental group $\Pi_1(\Sigma)$ (where $\Sigma$ is the base manifold of the supermembrane) into $SL(2,\mathbb{Z})$, the isotopy group of homotopic classes of symplectomorphisms (symplectomorphism group on 2-dimensions or equivalently area preserving diffeomorphisms is the local symmetry of the supermembrane in the Light Cone Gauge). The $SL(2,\mathbb{Z})$ group acts naturally on the first homology group of the fiber, which in our case corresponds to the target torus. The monodromy defines an automorphism on the fibers providing
the global structure of the geometrical setting.  We also show the existence of a new $Z_2$ symmetry
that plays the role of T-duality in the supermembrane interchanging the
winding and KK charges but leaving the Hamiltonian invariant, so
that the complete symmetry group in the ungauged supermembrane
corresponds to: $(SL(2,\mathbb{Z})_{\Sigma}\times
SL(2,\mathbb{Z})_{T^2})/Z_2$. T-duality becomes an exact symmetry of the
symplectic torus bundle description of the supermembrane by fixing its energy tension.

 In type IIB nine dimensional supergravities, there are
four inequivalent gaugings of $GL(2,\mathbb{R})$ global symmetry:
three of them are associated to the gauging of the
$SL(2,\mathbb{R})$ global symmetry: the parabolic, elliptic
and hyperbolic inequivalent classes and we find their respective
symplectic torus bundles. The fourth gauging corresponds to the
gauging of the trombone symmetry  associated to the $\mathbb{R}^+$
scalings. At quantum level the realization of this last gauging is
more involved since the scaling is not included in the arithmetic
subgroup $GL(2,\mathbb{Z})$. In \cite{cremmer} they provided a way
to realize this symmetry as a rigid symmetry, by studying a nonlinear realization of this symmetry
that was called active $SL(2,\mathbb{Z})$ symmetry. A way to
realize this scaling is by a nonlinear representation of $SL(2,\mathbb{Z})$. We show that this `symmetry' is present in the
ungauged supermembrane with central charges theory.  The symplectic
torus bundle associated to the gauging of this scaling symmetry is
constructed and it corresponds from the point of view of fibration
to a inequivalent class of symplectic torus bundles.  This proves the supermembrane
origin of the type IIB gauged supergravities. The monodromies
with type IIA origin are infered from the fact that T-duality invariance of the mass operator of the supermembrane with central charges.

The paper is structured in the following way: In section 2 we made a
summary of the results of  inequivalent classes of type IIB gauged
supergravities in 9D and its relation with the different
monodromies. In section 3 we summarize the construction of the
supermembrane with central charges, the two
$SL(2,\mathbb{Z})_{\Sigma}\times SL(2,\mathbb{Z})_{T^2}$ discrete
global symmetries.
 In section 4 we explain the sculpting
mechanism in which principle torus
fibration is deformed to acquire a monodromy of the fiber bundle. The corresponding action is gauged with
respect to the one already published in several works, see for example \cite{mor}. The new results are presented in sections 5,6,7, and 8. In section 5 we show the explicit global
construction of the gauged supermembrane with central charges, and the inequivalent classes of symplectic torus bundles associated to the the inequivalent classes of
monodromies. It is important to remark that for monodromies which include, elliptic, parabolic and hyperbolic classes there are torsion elements in the second cohomology group of the base manifold with coefficients in the module associated to the monodromy and this provides an extra restriction on the possible values of the charges of the theory. In section 6 we present the classification of the supermembrane theory formulated on the symplectic torus fibrations, and its relation to the different gaugings. We also discuss the residual symmetries of the theory after the gauge fixing. In section 7 we discuss
the fiber bundle construction for the supermembrane with the gauging of the trombone
symmetry.  The effect of the nonlinear representation of the
monodromy induces changes in the homology coefficients of the torus
of the fiber leading to inequivalent fibrations. 
In section 8 we show the existence of a new $Z_2$ symmetry
that plays the role of T-duality in the supermembrane.
 For other approaches to the supermembrane T-duality see \cite{aldabe1},\cite{aldabe2},\cite{russo}. In section 9 we
present our discussion and conclusions.
\section{Preliminars}

The gauged supergravities were firstly discovered by \cite{bedwn} by
compactifying the 11D supergravity on a $S^7$ a compact manifold
with nontrivial holonomy, soon after this result, the gauging
mechanism was also applied to theories with noncompact symmetry
groups in \cite{hullnoncompact}. The first paper of supergravity in nine dimensions containg a gauged sector was studied long time ago by \cite{gates}. Since then, the field has been very
active and it has been found a number of ways to obtain a consistent
deformation of a given maximal supergravity formulated in a target
space with $d<11$: by means of twisting in a Scherk-Schwarz
compactification (SS), through compactification on manifolds with
fluxes, noncommutative geometries etc.. For very nice reviews see
for example: \cite{7samtleben}, \cite{roest}.

In this section we will only review  aspects -all of them previously
found in the literature-, that are relevant for our constructions:
those in which  monodromy plays a fundamental role.
SS-compactifications appeared as a generalization of Kaluza-Klein
(KK)-reductions in which the fields are allowed to have a nontrivial
dependence on the compactified variables, but in such a way that the
truncation of the Langrangian in lower dimensions is still
consistent. SS-compactifications of supergravity may be expressed the
D-dimensional backgrounds in terms of principal fiber bundles over
circles with a twisting given by the monodromy \cite{hull-massive},\cite{pope}.
The background possesses  a group of global isometries $G$
associated to the compactification manifold over which it is
fibered. The principal fiber bundles of fiber $G$ have a monodromy
$\mathcal{M}(g)$ valued in the Lie algebra $g$ of the symmetry group
$G$. The invariant functional of the actions are expressed in terms
of the local sections of this bundle. The monodromy $\mathcal{M}(g)$
can be expressed in terms of a mass matrix $M$, as ${\cal M} (g)=
\exp M$. The maps in terms of the compactified variables $g(y)$ are
not periodic, but have a monodromy $g(y)= \exp (My)$
\cite{hull-massive}.

As explained in \cite{dh2} twisted compactification induces a SS-potential in the moduli space. For certain values of the moduli space it is
equivalent to introduce fluxes along the internal coordinates of the
compactified torus. In \cite{hull-townsend} it is conjectured that at
quantum level the global symmetry of the supergravity action breaks
to its arithmetic subgroup also called the U-duality group
$G(\mathbb{Z})$ . The quantization condition is imposed to preserve
the quantization of lattice of charges of the p-brane considered. At
quantum level all twisting must then belong to the $G(\mathbb{Z})$
duality group what implies also the restriction to quantized
parameters of mass matrix $M$. Indeed this condition was explored in
further detail in \cite{bgd} for the case of gauged
supergravities in 9D, where in addition to impose the elements of the
mass matrix to be integer, they have to satisfy in many cases, the
diophantine equation to guarantee that the monodromy lies in the
inequivalent classes of $SL(2,\mathbb{Z})$.

For the case of interest here, the type II gauged supergravities in
9D, the monodromies are associated to the $GL(2,\mathbb{R})= SL(2,\mathbb{R})\times
\mathbb{R}^{+}$ global symmetry group. In the $SL(2,\mathbb{R})$ sector,
there are three  inequivalent classes of theories, corresponding to
the hyperbolic, elliptic and parabolic $SL(2,\mathbb{R})$ conjugacy classes
and  represented by the monodromy matrices of the form
\cite{hull-massive} \bea \label{m1} {\mathcal M}_p=\begin{pmatrix}1 & k \\ 0 & 1 \end{pmatrix}, \qquad  {\mathcal M}_h=\begin{pmatrix} e^\gamma & 0 \\ 0
& e^{-\gamma} \end{pmatrix}, \qquad {\mathcal M}_e=\begin{pmatrix} \cos
\theta & \sin \theta \\ - \sin \theta & \cos \theta \end{pmatrix},
\qquad \eea
where each class is specified by the coupling constant ($k$, $\gamma$ or $\theta$).
In 9D the theory can also be described in terms of the mass matrix $M$
 with three parameters \cite{tp} \bea
M=\frac{1}{2}
\begin{pmatrix}
m^1 & m^2 +m^3 \\ m^2- m^3 & -m^1
\end{pmatrix}.
\eea

This mass matrix, as already explained in \cite{tp}, belongs to the Lie
algebra $sl(2,\mathbb{R})$ and transforms in the adjoint irreducible
representation. It is characterized by the vector
of mass $\stackrel{\rightarrow}{m}=( m_1,m_2,m_3)$.  At low energies the gauged supergravity is determined by the mass  matrix $M$ for a given monodromy  $\mathcal{M}$.

The field content of 9D II supergravity following the
notation of \cite{tp},\cite{bergshoeff}  is composed of a supervielbein $e_\mu{}^a$, three scalars
$\phi, \varphi, \chi,$ three gauge fields $(A_\mu, \{A^{(1)}_\mu,
A^{(2)}_\mu\}\equiv \vec A)$ two antisymmetric 2-forms
$\{B^{(1)}_{\mu\nu}, B^{(2)}_{\mu\nu}\}\equiv \vec B$, a three form
$C_{\mu\nu\rho}$ for the bosonic sector and in the fermionic side the
contribution is a spinor $\psi_\mu$ and two dilatinos $\lambda,
\tilde \lambda $ where the D=9 global $\Lambda =
    \left( \begin{array}{cc} a&b\\c&d \end{array} \right) \in SL(2,\mathbb{R})$ symmetry
acts in the ungauged theory in the following way:
\begin{align}
  {\tau} & \rightarrow \frac{a {\tau} +b}{c{\tau} +d} , \qquad
  \vec{{A}} \rightarrow \Lambda \vec{{A}}, \qquad
  \vec{{B}} \rightarrow \Lambda \vec{{B}},
\end{align} plus the fermionic transformations. One of the scalars $\varphi$ and
the three form $C$ remain invariant. As explained in
\cite{tp},\cite{bergshoeff} the gauge transformations correspond
to
\begin{align}
  A & \rightarrow A - d\lambda \qquad &
  \vec{B} & \rightarrow \vec{B} - \vec{A} d\lambda.
\end{align}

The massive deformations from the type IIB sector are labeled by four parameters $m=(m_i,m_4)\quad i=1,\dots,3$. Three of them $\vec{m} = (m_1,m_2,m_3)$ belong to the $SL(2,\mathbb{R})$ deformations and the last $m_4$ has its origin in the gauging of the scaling symmetry $\mathbb{R}^{+}$. The parameters of $m$
gauge a subgroup of the global symmetry $SL(2,\mathbb{R})$ and $\mathbb{R}^{+}$ respectively, with
parameter  $\Lambda =e^{\widetilde{M}\lambda}$ and gauge field transformations become modified as follows: \bea
  A \rightarrow A -  d \lambda  \qquad
  \vec{B} \rightarrow \Lambda
    ( \vec{B} - \vec{A} d\lambda ).
\eea
where we define $\widetilde{M}= (M,m_4)$, to group both type of deformations.
Following \cite{tp}, \cite{bergshoeff}, consider in first place the massive deformations associated to $\Lambda_{SL(2,\mathbb{R})}$ to the gauging of the
subgroup of $SL(2,\mathbb{R})$ with generator the mass matrix $M$
employed in the reduction. There are three distinct cases depending on the
value of $\vec {m}^2 = \tfrac14 (m_1{}^2 + m_2{}^2 -m_3{}^2)$
\cite{hull-massive,hull-gauged} characterizing the a set of three conjugacy
classes already shown in (\ref{m1}): $\mathbb{R},SO(1,1)^+, SO(2)$. Since we will make use of
them we will describe them shortly\footnote{To simplify the notation
we keep the one used in \cite{bergshoeff} and summarize their
results focusing only in the monodromy analysis.}. Each of the
subgroups is generated by a $SL(2,\mathbb{R})$ group element
$\Lambda$ with det\,$\Lambda = 1$. They are classified according to
their trace:
\begin{itemize}

\item The parabolic gauged supergravity is associated to the gauging
of the subgroup $\mathbb{R}$ with  parameter $\zeta$ generated
by
\begin{equation}
\Lambda_p =
 \left( \begin{array}{cc} 1&\zeta\\0&1 \end{array} \right)\, .
\end{equation}

The conjugacy class corresponds to matrices with $\vert Tr\Lambda_p\vert
=2$.

\item The hyperbolic gauged supergravity is associated to the gauging
of the subgroup $SO(1,1)^+$ with parameter $\gamma$
\begin{equation}
\Lambda_h =
  \left( \begin{array}{cc} e^{\gamma}&0\\
0&e^{-\gamma} \end{array} \right)\, .
\end{equation}
The conjugacy class is formed with matrices whose $\vert Tr\Lambda_h\vert
>2$

\item The elliptic gauged supergravity is associated to the gauging of the subgroup
 $SO(2)$ generated by elements $\Lambda_e$
of $SL(2,\mathbb{R})$ with parameter $\theta$,

\begin{equation}
 \Lambda_e =
\left( \begin{array}{cc} {\rm cos}\, \theta& {\rm sin}\, \theta
\\-{\rm sin}\, \theta& {\rm cos}\,\theta \end{array} \right)\, .
\end{equation}
 The elliptic conjugacy class correspond to matrices with $\vert Tr\Lambda_e\vert<2$.
\end{itemize}

The group $\mathbb{R}^+$ is a one-parameter conjugacy class. It
corresponds to the scalings that leave invariant the field
equations but scale globally the lagrangian. These symmetries where
called trombone by \cite{cremmer}. Its gauging was studied for example in
\cite{samtleben-trombone}, \cite{riccioni}. It
corresponds to the reduction with $m_{4}\ne 0;
m_1=m_2=m_3=0$. Following \cite{bergshoeff} the ${\mathbb
R}^+$-symmetry has been gauged with parameter $\Lambda_{\mathbb{R}^+} = e^{ m_{4}\lambda}$

As explained, in \cite{bergshoeff} the complete set of deformations
$\{m_i,m_{4}\}$ for the IIB reductions corresponds to \bea
\Lambda_{GL(2,R)}=\Lambda_{SL(2,R)}\Lambda_{\mathbb{R}^+}. \eea

At quantum level the realization of these symmetries $G$ is
proposed to be associated to their arithmetic subgroups $G(Z)$
\cite{hull-townsend}. The quantum realization of the trombone symmetry is more
involved. The problem at quantum level is the following: The group
$GL(2,\mathbb{R})$ should break to its arithmetic subgroup to guarantee the
quantization of the BPS charge lattice, however the set of matrices
$Mat(2,\mathbb{Z})$ whose determinant is an integer does not form a group
since the inverse of an integer is not necessarily an integer. (The
arithmetic subgroup of $GL(2,\mathbb{R})$ is the group
$GL(2,\mathbb{Z})=SL(2,\mathbb{Z})\times Z_2$, but it does fail in
incorporating the scalings). In \cite{cremmer} they found a proper
way to model out the scalings at quantum level by introducing
nonlinear representations of $SL(2,\mathbb{Z})$ that they called
active, to distinguish from those associated to the U-duality. This symmetry is
characterised by the fact that it acts on the lattice charge
transforming integer charges into integer charges by
$SL(2,\mathbb{Z})$ transformation but leaving the moduli fixed. This
is achieved by the use of a compensation transformation, that it is
applied once the U-duality transforms charges and moduli by the
linear $SL(2,\mathbb{Z})$, acting on the transformed moduli to get
it back to its original value.

\section{The Supermembrane with a topological condition}

In this section we will make a self-contained summary of the
construction of supermembrane with central charges due to a
topological condition. The hamiltonian of the $D=11$ Supermembrane
\cite{bst} may be defined in terms of maps $X^{M}$,
$M=0,\dots, 10$, from a base manifold $R\times \Sigma$, where
$\Sigma$ is a Riemann surface of genus $g$ onto a target manifold
which we will assume to be 11D Minkowski. Following \cite{hoppe},\cite{dwhn} one may now fix the Light
Cone Gauge, (LCG),
\bea X^{+}=T^{-2/3}P^{0+}\tau=-T^{-2/3}P^{0}_{-}\tau,
\quad P_{-}=P^{0}_{-}\sqrt{W}, \quad \Gamma^{+}\Psi=0 \eea
where $\sqrt{W}$ is a time independent density introduced in order to
preserve the density behavior of $P_{-}$.
$X^{-}, P_{+}$ are eliminated from the constraints and solve the fermionic second
class constraints in the usual way \cite{dwhn}.

The canonical reduced hamiltonian to the light-cone gauge
has the expression \cite{dwhn}

\begin{equation}\label{equ1}
  \mathcal{H}= \int_\Sigma  d\sigma^{2}\sqrt{W} \left(\frac{1}{2}
\left(\frac{P_M}{\sqrt{W}}\right)^2 +\frac{1}{4} \{X^M,X^N\}^2-
\overline{\Psi}\Gamma_{-}\Gamma_{M}\{X^M,\Psi\}\right)
\end{equation}
subject to the constraints \begin{equation}  \label{e2}
\phi_{1}:=d(\frac{P_{M}}{\sqrt{W}}dX^{M} -\overline{\Psi}\Gamma_{-}d\Psi)=0 \end{equation} and
\begin{equation} \label{e3}
\phi_{2}:=
   \oint_{C_{s}}(\frac{P_M}{\sqrt{W}}dX^M -\overline{\Psi}\Gamma_{-}d\Psi)= 0,
\end{equation}
where the range of $M$ is now $M=1,\dots,9$ corresponding to the
transverse coordinates in the light-cone gauge, $C_{s}$,
$s=1,2$ is a basis of  1-dimensional homology on $\Sigma$,
 \be \label{e4}\{X^{M}, X^{N}\}= \frac{\epsilon
^{ab}}{\sqrt{W(\sigma)}}\partial_{a}X^{M}\partial_{b}X^{N}. \ee
$a,b=1,2$ and $\sigma^{a}$ are local coordinates over $\Sigma$.
 $\phi_{1}$ and $\phi_{2}$ are generators of area preserving diffeomorphisms, see \cite{dwmn}. That is
\be \sigma\to\sigma^{'}\quad\to\quad W^{'}(\sigma)=
W(\sigma).\nonumber \ee When the target manifold is simply connected
$dX^{M}$ are exact one-forms.

We consider now the compactified Supermembrane embedded on a target space $M_9\times T^2$  where
$T^2$ is a flat torus defined in terms of a lattice $\mathcal{L}$ on
the complex plane $C$:
 \bea\label{3}
 \mathcal{L}: z\to z+2\pi R(l+m\tau),
 \eea
where $m,l$ are integers, R is real and represent the radius, $R>0$, and $\tau$ a
complex moduli $\tau=Re\tau +iIm\tau$, $Im\tau>0$, $T^2$ is defined
by ${C} / {\mathcal{L}}$. $\tau$ is the complex coordinate of the
Teichmuller space for $g=1$, that is the upper half plane. The
Teichmuller space is a covering of the moduli space of Riemann
surfaces, it is a $2g-1$ complex analytic simply connected manifold
for genus $g$ Riemann surfaces.

The conformally equivalent tori are identified by the parameter
$\tau$ modulo the Teichmuller modular group, which in the case $g=1$
is $SL(2,\mathbb{Z})$.   It acts on
the Teichmuller space through a Mobius transformation and it has a
natural action on the homology group $H_1(T^2)$.

We consider
maps $X^m,X^r$  from $M_9\times T^2$ to the target space , with $r=1,2;
 m=3,\dots,9$ where $X^m$ are single valued maps onto the Minkowski sector of the target
space while $X^r$ maps onto the $T^2$ compact sector of the
target. The winding condition corresponds to

\bea\begin{aligned}
&\oint_{\mathcal{C}_{s}}dX=2\pi R (l_{s}+m_{s}\tau)\\
&\oint_{\mathcal{C}_{s}}dX^m=0 \end{aligned}\eea where  $dX= dX^1+idX^2$ and $l_{s},
m_{s}, s=1,2$, are integers.
We denote
$d\widehat{X}^{r}$, $r=1,2$, the normalized harmonic one-forms with respect to
$\mathcal{C}_{s}$, $s=1,2$, a canonical basis of homology on
$\Sigma$: \bea \label {e7}
\oint_{\mathcal{C}_{s}}d\widehat{X}^{r}=\delta^{r}_{s}.\eea

We now impose a topological
restriction on the winding maps \cite{mrt}: the irreducible winding constraint,
 \bea\label{central}
 \int_{\Sigma}dX^r\wedge dX^s=n\epsilon^{rs}Area(T^2)\quad r,s=1,2
 \eea

Using $Area(T^2)=(2\pi R)^2 Im\tau$, condition (\ref{central}) implies that the winding matrix $\mathbb{W}= \begin{pmatrix} l_{1}& l_{2}\\
                       m_{1} & m_{2}\end{pmatrix}$ has  $det \mathbb{W}=n\ne 0$. That is, all integers $l_s,m_s, s=1,2$ are admissible provided
$det \mathbb{W}= n$ when $n$ is assumed to be different from
zero. $\epsilon^{rs}$ is the symplectic antisymmetric tensor
associated to the symplectic 2-form on the flat torus $T^2$. In the
case under consideration $\epsilon^{rs}$ is the Levi Civita
antisymmetric symbol.

We may decompose the closed one-forms $dX^{r}$ into
 \bea
dX^{r}=M_{s}^{r}d\widehat{X}^{s}+dA^{r}\quad r=1,2 \eea where
$d\widehat{X}^{s}, s=1,2$ is the basis of harmonic one-forms we have
already introduced, $dA^{r}$ are exact one-forms and $M_{s}^{r}$ are
constant coefficients.  This condition is satisfied provided
\bea M_{s}^{1}+iM_{s}^{2}=2\pi R (l_{s}+m_{s}\tau) \eea
Consequently, the most general expression for the maps $X^{r}$,
$ r=1,2$, is \bea\label{9} dX=2\pi R (l_{s}+m_{s}
\tau)d\widehat{X}^{s}+ dA, \eea $l_{s}, m_{s}$, $s=1,2$, arbitrary
integers.

 An important point implied by the assumption
$n\ne 0$ is that the cohomology class in $H^2(\Sigma,Z)$ is
non-trivial. It also implies that at global level the theory is
described by an action formulated on a principal torus bundle
over $\Sigma$. There exists a infinite set of possible gauge
connections associated to it.

The topological condition (\ref{central}) does not change the field
equations of the hamiltonian (\ref{equ1}). In fact, any variation of $I^{rs}$ under a change $\delta X^{r}$,
single valued over $\Sigma$, is identically zero. In addition to the
field equations obtained from (\ref{equ1}), the classical
configurations must satisfy the condition (\ref{central}). It is
only a topological restriction on the original set of classical
solutions of the field equations. In the quantum theory the space of
physical configurations is also restricted by the condition
(\ref{central}). There is a compatible election for $W$ on the
geometrical picture we have defined.  We
define \bea \label{8}
\sqrt{W}=\frac{1}{2}\epsilon_{rs}\partial_{a}\widehat{X}^{r}\partial_{b}\widehat{X}^{s}\epsilon^{ab},
\eea it is a regular density globally defined over $\Sigma$. It is
invariant under a change of the canonical basis of homology.

The physical hamiltonian in the LCG is given by \begin{equation}\label{5}
\begin{aligned}
&\mathcal{H}= \int_{\Sigma}T^{-2/3}\sqrt{W}\left[\frac{1}{2}(\frac{P_{m}}{\sqrt{W}})^{2}+
\frac{1}{2}(\frac{P_{r}}{\sqrt{W}})^{2}
+\frac{T^{2}}{2}\{X^{r},X^{m}\}^{2}
+\frac{T^{2}}{4}\{X^{r},X^{s}\}^{2}\right]\\ \nonumber &+\int_{\Sigma}T^{-2/3}\sqrt{W}\left[\frac{T^{2}}{4}\{X^{m},X^{n}\}^{2}- \overline{\Psi}\Gamma_{-} \Gamma_{m} \{X^m,\Psi\}- \overline{\Psi}\Gamma_{-} \Gamma_{r} \{X^r,\Psi\}\right] \end{aligned}\end{equation} subject to the constraints \bea\label{6}
 d(P_{r} dX^{r}+P_{m}dX^{m}-\overline{\Psi}\Gamma_{-}d\Psi)=0\eea\begin{equation} \label{6a}
\oint_{\mathcal{C}_{s}}(P_{r} dX^{r}+P_{m}dX^{m}-\overline{\Psi}\Gamma_{-}d\Psi)=0 \end{equation} and the
global restriction (\ref{central}). $\mathcal{C}_{s}$ is a canonical
basis of homology on $\Sigma$. This is the case with trivial monodromy and hence without the gauging of the $SL(2,\mathbb{Z})$ symmetries described below. It is a symplectic gauge theory on a given isotopy class of symplectomorphisms.

The Mass operator of the supermembrane with central charges and KK modes found in \cite{sl2z} is
\bea\label{28} Mass^{2}=T^{2}((2\pi R)^{2}n Im \tau)^{2}+
\frac{1}{R^{2}}((m_{1}^{2}+(\frac{m\vert q\tau-p\vert}{R Im\tau})+ T^{2/3}H\eea where the $H$ is defined in terms of the above hamiltonian $\mathcal{H}$ once the winding contribution has been extracted $H= \mathcal{H}-
T^{-2/3}\int_{\Sigma}\sqrt{W}\frac{T^{2}}{4}\{X^{r}_{h},X^{s}_{h}\}^{2}$.
\subsection{The $SL(2,\mathbb{Z})$ Symmetries of the supermembrane with central charges}

The supermembrane is invariant under are preserving diffeomorphisms on the base manifold. This symmetry is realized by the first class constraints on the theory. This is a gauge symmetry associated to a trivial principle bundle with structure group the symplectomorphisms homotopic to the identity. Besides this standard symmetry of the supermembrane, when the theory is restricted by the central charge condition (the irreducible winding condition), the theory is invariant under two $SL(2,\mathbb{Z})$ symmetries. One of them acting on the homology basis of the base manifold $\Sigma$, a two-torus. This $SL(2,\mathbb{Z})$ realizes the modular transformations\footnote{In particular the supermembrane with central charges is invariant under the conformal maps homotopic to the identity.} on the upper-half plane. The other $SL(2,\mathbb{Z})$ acts on the target space, on the moduli of the target torus: the complex $\tau$ and $R$ parameters of the target torus. On $\tau$ acts as a Moebius transformations, however since the transformation of $R$ is nontrivial, the equivalence classes of tori under this transformation are not conformally equivalent.
 Using these two $SL(2,\mathbb{Z})$ symmetries, it can be seen \cite{sl2z} that the mass
 contribution of the stringy states in the supermembrane with central charges exactly agree with
 the perturbative mass spectrum of $(p,q)$ $IIB$ and $IIA$
 superstring. Let us discuss it in more detail:

\subsubsection{$SL(2,\mathbb{Z}$) of the Riemann surface}

The supermembrane with central charges is invariant under area preserving diffeomorphisms
 homotopic to the identity. Those are
diffeomorphisms which preserve $d\widehat{X}^{r}, r=1,2$, the
harmonic basis of one-forms. $W$ is then invariant: \bea
W^{'}(\sigma)=W(\sigma). \eea Moreover the supermembrane with central charges is invariant under
diffeomorphisms changing the homology basis, and consequently the
normalized harmonic one-forms, by a modular transformation on the
Teichm\"uller space of the base torus $\Sigma$. In fact, if
\bea\label{12} d\widehat{X}^{'r}(\sigma)=S^{r}_{s}
d\widehat{X}^{s}(\sigma) \eea provided \bea
\epsilon_{rs}S_{t}^{r}S_{u}^{s}=\epsilon_{tu} \eea that is $S\in
Sp(2,Z)\equiv SL(2,\mathbb{Z})$. We then conclude that the supermembrane with central charge, has
an additional symmetry with respect to the compactified $D=11$ Supermembrane without the topological irreducibility condition. All
conformal transformations on $\Sigma$ are symmetries of the supermembrane with central charges
\cite{gmr}-\cite{bgmr3}. We
notice that under (\ref{12}) \bea \label{15}dX\to 2\pi R
(l^{'}_{s}+m^{'}_{s}\tau)d\widehat{X}^{'s}+dA^{'} \eea where
$A^{'}(\sigma^{'})=A(\sigma)$ is the transformation law of a scalar.
Defining the winding matrix as  $\mathbb{W}=\begin{pmatrix} l_{1}& l_{2}\\
               m_{1} & m_{2} \end{pmatrix}$, then

\begin{equation}\label{16}
\mathbb{W}\to \mathbb{W} S^{-1}
\end{equation}

\subsubsection{The U-duality invariance}

The supermembrane with central charges is also invariant under the
following transformation on the target torus $T^{2}$:
\bea \label{17}\tau &\to& \frac{a\tau+b}{c\tau+d}\\
\nonumber  R& \to & R|c \tau +d| \\ \nonumber  A &\to& A
e^{i\varphi}
\\
\nonumber  \mathbb{W}&\to& \begin{pmatrix} a & -b\\
                        -c & d
 \end{pmatrix}\mathbb{W}\eea

where $c\tau+d= |c\tau+d|e^{-i\varphi}$ and $\Lambda=\begin{pmatrix} a & b\\
                        c & d
 \end{pmatrix}\in Sp(2,Z)$. As shown in \cite{sl2z} the hamiltonian density of the supermembrane with central charges is then
 invariant under (\ref{17}). The $SL(2,\mathbb{Z})$ matrix now acts from the
 left of the matrix $\mathbb{W}$.

The two actions from the left and from the right by $SL(2,\mathbb{Z})$
matrices are not equivalent, they are complementary. The following
remarks are valid. The general expression for the $dX$ maps is then
 \bea
dX=dX_{h}+dA
 \eea

The harmonic part of $dX$,
\bea\label{nt11}
dX_{h}=2\pi R [(m_1\tau+l_1) d\widehat{X}^{1}+(m_2\tau +l_2) d\widehat{X}^{2}].
\eea

$X_h$ is a minimal immersion from $\Sigma$ to $T^{2}$ on the target,
moreover it is directly related to a holomorphic immersion of
$\Sigma$ onto $T^{2}$. The extension of the theory of supermembranes
restricted by the topological constraint to more general compact
sectors in the target space is directly related to the existence of
those holomorphic immersions.

\section{The Sculpting Mechanism for Gauging Theories}

In this section we summarize the results of the paper \cite{sculpting}.
The mechanism of gauging proposed there consists in a specific
change in the global description of a theory in terms of fibration,
it is called \textit{sculpting mechanism}.  It consists in a
deformation of the homotopy-type of the complete fibration
preserving the homotopy-type of the base and the fiber. We will
restrict here to the application of this mechanism to the
supermembrane. Taking as the un-gauged theory the compactified
supermembrane on a 2-torus. It corresponds to a invariant functional
(action) over a Riemann base manifold whose fiber is the tangent
space: $T^2\times M_9$ for simplicity. The topologically nontrivial
part of the fiber corresponds to the torus manifold associated to
the tangent space. The global formulation of the un-gauged theory is
a trivial torus bundle over a base manifold that for simplicity we
also choose to be homotopically a torus.\newline

The change of the total fiber bundle can be viewed in terms of two
main steps: the first one is due to the introduction of a
topological condition that we will explain below (the central
charge) by which the trivial torus bundle is deformed into a
principal  bundle. On the physical side, it can be seen a
restrictions on the maps allowed in the compactified target space.
Secondly, the process of extracting the gauge field from the closed
form in a consistent way implies the modification of the principal
torus bundle in a symplectic torus bundle with monodromy.  The total
fiber bundle may or may not be symplectic according to the fact that
the monodromy is given by the torsion class associated to the MCG of
the $\Pi_1(\Sigma)$ base manifold. The resulting supermembrane is
therefore, gauged in this new \textit{sculpting} sense and it
corresponds geometrically to a supermembrane minimally immersed in
the target space. As a result of this procedure, the global symmetry
of the un-gauged theory is partially broken to a subgroup $H\in G$.
A new gauge symmetry $A$ appears due to the global as a restriction
of the diffeomorphism invariance gauge symmetry of the compact base
manifold by the discrete symmetry subgroup $\Gamma$ associated to
the monodromy representation $\rho$ of the harmonic forms. This
symmetry gets promoted to a connection by the action of the
principal fiber bundle to which the symplectic torus bundle is
associated to.

The change in the homotopy-type of the complete manifold is produced
by extracting properly the gauge connection from the closed 1-forms.
The supermembrane with central charges as a global manifold
corresponds to a symplectic torus bundle with nontrivial
monodrodromy $\rho$. The Cohomology of the torus bundle change in
this case in the following way \bea \mathbb{I}\leadsto
H^2(\Sigma,\mathbb{Z})\leadsto H^2(\Sigma,\mathbb{Z}_{\rho}) \eea

being $\rho$  a representation of the large diffeomorphims group of
the base manifold. Notice that arrows do not imply a spectral sequence. Closely following \cite{sculpting} we just emphasize
the three main steps needed to produce the \textit{sculpting}
deformation of the fiber bundle: I The first step is to impose the
central charge condition which represents a obstruction to the
triviality called that produces a twist in the fibration generating
a principal fiber bundle whose cohomology is
$H^2(\Sigma,\mathbb{Z})$. The lagrangian of the undeformed fiber has
the following symmetries: a gauge symmetry  $DPA_0(\Sigma_1^2)$,
target space susy $N=2$, a discrete global symmetry $G\equiv
Sp(2,Z)$ associated to the wrapping condition of the embedded maps
$\Sigma_1^2\to T^2$: There exists a infinite set of connections that
can be attached to the principal bundle . The winding condition
defines closed 1-forms $dX_r$ that admit a Hodge decomposition in
terms of harmonic one-forms $d\widehat{X}_r$ and a exact one-form
$dA_r$:
 \bea
 dX_r=P_r^s d\widehat{X}_s+dA_r
 \eea

the matrix $P_r^s$ is associated to the  4 global degrees of freedom
associated to the winding condition, whose coefficients depend on
time. In presence of the central charge condition, the matrix
$\mathcal{P}_r^s$ becomes constant and non-degenerate, (we are
freezing the wrapping).

The harmonic one-forms due to the wrapping condition have a global
Sp(2,Z) symmetry of the mapping class group. As a consequence of the
nontrivial fibration now \bea P_r^s= M_{r}^s= 2\pi R^r S_r^s\quad
with \quad S_r^s\in SL(2,\mathbb{Z})\eea

Once a fixed basis $\{d\hat{X}^s\}$, is chosen , the decomposition
is unique, and $P_r^s$ is fixed (for example to $\delta^r_s$) there
is a partial fixing of the symmetry that breaks the original global
symmetry to a residual one, that leaves a global invariance under
the subgroup that will be related to the monodromies associated to
the gauging. \newline The next step is to \textit{extract} a
one-form connection to the nontrivial fiber bundle. We define a
symplectic connection $A$ preserving the structure of the fiber
under holonomies. To this end, first we define a rotated derivative
associated to the Weyl bundle \cite{mrsymp}:
  \bea
  D_r\bullet=( 2\pi R^rl^r)\theta_l^r\frac{\epsilon^{ab}}{\sqrt{W(\sigma)}}\partial_a\widehat{X}^l(\sigma)\partial_b \bullet
  \eea
 with $\theta\in SL(2,\mathbb{Z})$ which depends on the monodromy $\rho$.

In 2-dim the area preserving diffeomorphisms are the same as the
symplectomorphisms. The third relevant choice is the election for
$W$ on the geometrical picture we have defined. We define \bea
\label{8}
\sqrt{W}=\frac{1}{2}\epsilon_{rs}\partial_{a}\widehat{X}^{r}\partial_{b}\widehat{X}^{s}\epsilon^{ab},
\eea it is a regular density globally defined over $\Sigma$. It is
invariant under a change of the canonical basis of homology.

The matrix $\theta$ carries the information of the discrete global
symmetry residual associated to transition functions of the patching
of the different charts in the compact base manifold for a fixed
base of harmonic forms. It plays a analogous role to the embedding
tensor in the Noether gauging of supergravities theories. Let us
signal that here the place where the discrete global symmetries
appear together with the derivative operator instead of appearing
besides the gauge field since its origin its topological associated
to the p-brane base manifold compact surface.

The definition of this rotated derivative, we are performing an
\textit{extension of the covariant derivative definition}, in which
the associated bundle has a nontrivial monodromy from the $\pi_1
(\Sigma)$ on the homology of the fiber $H_1 (T^2)$ . The related
derivative fixes a scale in the theory and breaks the former $H=
Sp(2,\mathbb{Z})$ theory to a subgroup $\Gamma\in Sp(2,\mathbb{Z})$ by specifying the
integers of $S_r^s$. Fixing $R^r$ also fixes the Kahler and complex
structure geometrical moduli.

The symplectic covariant derivative \cite{mrsymp}, is then:
  \bea
  \mathcal{D}_r\bullet=D_r\bullet+\{A_r,\bullet\}
  \eea
  and then the connection transform with the symplectomorphism like:
  \bea
  \delta_{\epsilon}A=\mathcal{D}_r\epsilon
  \eea

The \textit{sculpted} fiber bundle is a symplectic torus bundle with
cohomology $H^2(\Sigma,\mathbb{Z}_{\rho})$.

This symplectic form is one in particular different to the canonical
one associated to the flat torus $t^2$ taken as a starting point in
the compactified supermembrane case associated to the trivial torus
bundle. This means that the nontrivial fibration implies a
deformation in the base manifold, indeed the isometry group closely
related to the harmonic group of symmetry is not the associated to a
flat torus.Since a Riemann manifold has three compatible structures
$g_{ab},J,\Lambda_{ab}$ the metric is associated to the harmonic
one-forms that preserve the fiber associated to the MR-monopoles \cite{mr},
the induced symplectomorphism do not lie in the same conformal class
of the flat torus. There is a compatible election for $W$ on the
geometrical picture we have defined. We consider the $2g$
dimensional space of harmonic one-forms on $\Sigma$. We denote
$dX^{r}$, $r=1,2$, the normalized harmonic one-forms with respect to
$\mathcal{C}_{s}$, $s=1,2$, a canonical basis of homology on
$\Sigma$: \bea \label {7}
\oint_{\mathcal{C}_{s}}d\widehat{X}^{r}=\delta^{r}_{s}. \eea We
define \bea \label{8}
\sqrt{W}=\frac{1}{2}\epsilon_{rs}\partial_{a}\widehat{X}^{r}\partial_{b}\widehat{X}^{s}\epsilon^{ab},
\eea it is a regular density globally defined over $\Sigma$. It is
invariant under a change of the canonical basis of homology.

It also implies that there is an $U(1)$ nontrivial principle bundle
over $\Sigma$ and a connection on it whose curvature is given by
$d\widehat{X}^r\wedge d\widehat{X}^s$. This $U(1)$ nontrivial
principal fiber bundle are associated to the presence of monopoles
on the worldvolume of the supermembrane explicitly discussed in
\cite{mr}.


After replacing this expression in the hamiltonian (\ref{equ1}) one
obtain the gauged supermembrane in this new \textit{sculpting} sense gauging the $SL(2,\mathbb{Z})$ that is the
hamiltonian of the supermembrane with central charges \cite{mor}:

\bea\label{e9}
 \begin{aligned}
H&=\int_{\Sigma}\sqrt{W}d\sigma^{1}\wedge
d\sigma^{2}[\frac{1}{2}(\frac{P_{m}}{\sqrt{W}})^{2}+\frac{1}{2}
(\frac{P^{r}}{\sqrt{W}})^{2}+
\frac{1}{4}\{X^{m},X^{n}\}^{2}+\frac{1}{2}(\mathcal{D}_{r}X^{m})^{2}+\frac{1}{4}(\mathcal{F}_{rs})^{2}
\\ \nonumber &+(n^2 \textrm{Area}_{T^2}^2)+ \int_{\Sigma}\sqrt{W}\Lambda
(\mathcal{D}_{r}(\frac{P_{r}}{\sqrt{W}})+\{X^{m},\frac{P_{m}}{\sqrt{W}}\})]\\& + \int_{\Sigma}
\sqrt{W} [- \overline{\Psi}\Gamma_{-} \Gamma_{r}
\mathcal{D}_{r}\Psi -\overline{\Psi}\Gamma_{-}
\Gamma_{m}\{X^{m},\Psi\}-
 \Lambda \{ \overline{\Psi}\Gamma_{-},\Psi\}]
 \end{aligned}
 \eea
where  $\mathcal{D}_r X^{m}=D_{r}X^{m} +\{A_{r},X^{m}\}$,
$\mathcal{F}_{rs}=D_{r}A_s-D_{s }A_r+ \{A_r,A_s\}$, \\
 $D_{r}=2\pi l_r\theta_{r}^{l}
R_{r}\frac{\epsilon^{ab}}{\sqrt{W}}\partial_{a}\widehat{X}^{l}\partial_{b}$
and $P_{m}$ and $P_{r}$ are the conjugate momenta to $X^{m}$ and
$A_{r}$ respectively. $\mathcal{D}_{r}$ and $\mathcal{F}_{rs}$ are
the covariant derivative and curvature of a symplectic
noncommutative theory \cite{mrsymp}, constructed from
the symplectic structure $\frac{\epsilon^{ab}}{\sqrt{W}}$ introduced
by the central charge. The last term represents its
supersymmetric extension in terms of Majorana spinors. $\Lambda$ are the lagrange multiplier associated to the constrains. The physical
degrees of the theory are the $X^{m}, A_{r}, \Psi_{\alpha}$ they are
single valued fields on $\Sigma$.\newline

\section{The Supermembrane as a Symplectic Torus Bundle with Monodromy in $SL(2,\mathbb{Z})$.}

In this section we develop the global construction found in
\cite{gmmpr}, characterizing in deeper detail its connection with
the $SL(2,\mathbb{Z})$ gaugings in supergravity in 9D.\newline

We consider in this section the global structure of the supermembrane in the Light Cone Gauge
when the fields $X,\Psi$ are sections and $A$ is a symplectic
connection on a nontrivial symplectic torus bundle. A symplectic
torus bundle $\xi$ is a smooth fiber bundle $F\to
E\stackrel{\pi}{\rightarrow}\Sigma$  whose structure group $G$ is
the group of symplectomorphisms preserving a symplectic two-form on
the fiber $F$. $\Sigma$ is the base manifold which we consider to be
a closed, compact Riemann surface modeling the spacial piece of the
foliation of the supermembrane worldvolume, and $E$ is the total
space.  We will take the fiber as the target-space manifold
$M_9\times T^2$ consider in section 3, as in
\cite{sculpting},\cite{gmmpr}. The only topologically nontrivial
part corresponds to the $T^2$, so from now on, we will only refer to
this part that is the one that characterizes the fiber bundle. We
consider in particular  $\Sigma$, as already explained, a genus
$g=1$ surface with a non-flat induced metric. We remark that when
$g>1$, the first homotopy group $\Pi_1(\Sigma)$ is non-abelian
allowing the construction of symplectic torus bundles with
non-abelian monodromies. In this paper we will restrict to the
abelian case only.
\newline

On $T^2$, a flat torus, we consider the canonical symplectic 2-form.
Its pullback, using the harmonic maps from the base manifold to
$T^2$, defines the symplectic 2-form $\omega$ on $\Sigma$. In terms
of a harmonic basis of one-forms $d\widehat{X}^r$, $r=1,2$ in the
notation of Section 3: $\omega=[(2\pi R)^2 n Im\tau]
\epsilon_{rs}d\widehat{X}^r\wedge d \widehat{X}^s$. The
symplectomorphisms\footnote{ On a 2-dimensional surface
symplectomorphisms and area preserving diffeomorphisms define
the same group.} on $\Sigma$ homotopic to the identity are generated
by the first class constraints (\ref{e2}), (\ref{e3}). Moreover, the
symplectomorphisms preserving $\omega$ define isotopic classes.
These classes form a group $\Pi_0 (G)$ where $G$ is the group of all symplectomorphisms. In the case we are
considering, where the fiber is $T^2$, $\Pi_0(G)$ is isomorphic to
$SL(2,\mathbb{Z})$.  The action of $G$ on the fiber $T^2$ produces
an action on the homology and cohomology of $T^2$. This action
reduces to an action of $\Pi_0 (G)$, since on a given isotopy class
two symplectomorphisms are connected by a continuous path within the
class, and hence one cannot change the element of the homology or
cohomology group. The action of $G$ on the fiber over a point $x\in
\Sigma$ when one goes around an element of $\Pi_1 (\Sigma)$ defines a
homomorphism
 \bea
 \Pi_1(\Sigma)\to\Pi_0(G)\approx SL(2,\mathbb{Z})
 \eea

which may be called the monodromy of the symplectic torus bundle\footnote{It would be
interesting to see if there is a relation (if any) with a  construction on
torus bundles with monodromy that has recently appeared \cite{baraglia}.}.
The monodromy may be trivial or not, but even when it is trivial,
the symplectic torus bundle can be nontrivial. In fact, one could
have a nontrivial transition within the symplectomorphisms on a
isotopy class. If the monodromy is trivial, the symplectic torus
bundle is trivial if and only if there exists a global section.
When $\Sigma$ is a 2-torus, as we are considering,  $\Pi_1 (\Sigma)$
is abelian and the homomorphism defines a representation $\rho :
\Pi_1 (\Sigma)\to SL(2,\mathbb{Z})$, realized in terms of an abelian
subgroup of $SL(2,\mathbb{Z})$. It naturally acts on $H_1 (T^2)$ the
first homology group on $T^2$. This provides to $H_1 (T^2)$ the
structure of a $Z[\pi_1 (\Sigma)]$-module which may be denoted
$Z_{\rho}^2$. Given $\rho$ there is a bijective correspondence
between the equivalence classes of symplectic torus bundles with
base $\Sigma$ and $Z_{\rho}^2-$module, and the elements of
$H^2(\Sigma,Z_{\rho}^2)$, the second cohomology group of $\Sigma$
with coefficients $Z_{\rho}^2$ \cite{khan}. Following \cite {khan}:
the element of $H^2 (\Sigma,Z_{\rho}^2)$ is called the cohomology
class of the symplectic torus bundle and it is denoted $C(E)$.
$C(E)=0$ if and only if there exists a global section on $E$. If
$\rho$ is trivial, $C(E)=0$ if and only if $E$ is trivial.

The supermembrane theory with nontrivial central charge has
$C(E)\ne 0$ and hence $E$ is always nontrivial. The
supermembrane on a eleven dimensional Minkowski target space
\cite{dwhn} was formulated on a trivial symplectic bundle, as well
as the supermembrane on a compactified space in
\cite{dwpp}. The $C(E)\ne 0$ condition is the relevant
condition which ensures a discrete spectrum of supermembrane with
nontrivial central charges \cite{gmr}-\cite{bgmr3}.  In the case of a
trivial symplectic torus bundle the spectrum spectrum of the
supermembranes was proven to be continuous from
$[0,\infty)$\cite{dwln}. There is a third case , which has not been
discussed in the literature: $C(E)=0$ but a nontrivial monodromy.
The analysis of the spectrum of a supermembrane on such a symplectic
torus bundle could render a supermembrane theory with discrete
spectrum on the $C(E)=0$ sector, which is excluded by the
supermembrane with the nontrivial central charges. This important
point will be analyzed elsewhere.

The second cohomology $H^2(\Sigma,Z_{\rho}^2)$ may be equal to $Z$,
as in the case of the representation
 \bea\label{t1}
\rho(\alpha,\beta)=\begin{pmatrix} 1 & \alpha \\ 0 & 1
\end{pmatrix},\quad \textit{or}\quad
\rho(\alpha,\beta)=\begin{pmatrix} 1 & \beta \\ 0 & 1 \end{pmatrix}
\eea where $(\alpha,\beta)$ denotes the element of
$\Pi_{1}(\Sigma)$. But it may also have a finite number of elements
as in the case of \cite{khan}, \bea\label{t2}
\rho(\alpha,\beta)=\begin{pmatrix} -2mn+1 & 2mn^2+n \\ -m & mn+1
\end{pmatrix}^{(\alpha+\beta)} \eea

where the integers $m,n>0$. In this case
$H^2(\Sigma,Z_{\rho}^2)=Z_m\oplus Z_n$. The number of inequivalent
symplectic torus bundles is, in this case, $mn$. Hence given $\rho$
the number of inequivalent symplectic torus bundles is in general
not in the correspondence with $Z$ as one could in principle think.
This remark has relevant consequences in the analysis of the symmetry groups
associated to the theory at quantum level. From a geometrical point
there is a qualitative difference between the symplectic torus
bundle associated with the representations (\ref{t1}) and
(\ref{t2}). A theorem in \cite{khan} ensures the existence of
symplectic 2-form on $E$ which reduces to the the symplectic
2-form on each fiber if and only if the element $H^2 (\Sigma,
Z_{\rho}^2)$ associated to $E$ is a torsion element. In case
(\ref{t2}) all elements  are torsion while in case (\ref{t1}) only
$C(E)=0$, which is excluded if we consider a supermembrane with
nontrivial central charge. Let us now consider the transformation
law of the fields describing the supermembrane with nontrivial
central charge. In section 3, we showed the transformation law under a
rigid $SL(2,\mathbb{Z})$ transformations. There are two
$SL(2,\mathbb{Z})$ invariances, one associated to the basis $\Sigma$
and one to the moduli on the target space. We now consider a
supermembrane on a symplectic torus bundle with monodromy $\rho
(\alpha,\beta)$. Under a rigid $SL(2,\mathbb{Z})$ on the target the
symplectic connection $A(x)$ transforms with a global factor
$e^{i\varphi}$ where $e^{-i\varphi}=\frac{c\tau+d}{\vert
c\tau+d\vert}$ and $\Lambda=\begin{pmatrix} a& b \\ c&
d\end{pmatrix}\in SL(2,\mathbb{Z})$ acts on the moduli and winding matrix as already stated. On the
symplectic torus bundle with monodromy $\rho(\alpha,\beta)$, $A(x)$
transforms with a phase factor $e^{i\varphi_{\rho}}$ with
$\varphi_{\rho}\equiv \varphi(\rho(\alpha,\beta))$ but now
$\Lambda\equiv \rho(\alpha,\beta)$. That is $a,b,c,d$ are integers
which depend on $(\alpha,\beta)$. For example, if we consider
$\alpha=\beta=0$ corresponding to a trivial element of
$\Pi_1(\Sigma)$ then  $\varphi=0$, while if $(\alpha,\beta)\ne (0,0)$ then $\varphi$ can be different from zero , for
example in case (\ref{t2}). If we write $A(x)=\vert A(x)\vert
e^{i\lambda(x)}$ then associated to $(\alpha,\beta)\in
\Pi_1(\Sigma)$ we have $A(x)=\vert A(x)\vert
e^{i\lambda(x)+\varphi_\rho}$. We then have,

 \bea
 d\left( \vert A(x)\vert e^{i\lambda(x)+e^{i\varphi_\rho}}\right)=dA(x)e^{i\varphi_\rho}.\eea

 In order to take into account the phase factor $e^{i\varphi_{\rho}}$ we may multiply the
 symplectic covariant derivative in the formulation by this phase factor and leave $A(x)$
 as a single-valued one-form connection. In the hamiltonian of section 3, the
phase factor $e^{i\varphi_{\rho}}$ is canceled by its
complex conjugate contribution consequently, the hamiltonian is
well-defined on a symplectic torus bundle with nontrivial monodromy.
Another important aspect of the supermembrane formulated on a
symplectic torus bundle with monodromy is that the $(p,q)$
Kaluza-Klein charges in the mass squared formula take value on the
$Z_{\rho}^2$-module. In fact, the $(p,q)$ charges are naturally
associated to the element of $H_1 (T^2)$. We then have a
nice geometrical interpretation: The KK charges are associated to
the homology of $T^2$ on the target, while the winding is associated
to the cohomology on the base $\Sigma$. In \cite{gmmpr} we proved
that the hamiltonian together with the constrains are invariant
under the action of $ SL(2,\mathbb{Z})$ on the homology group
$H_1(T^2)$ of the fibre 2-torus $T^2$. So that the supermembrane
with central charges may be formulated in terms of sections of
symplectic torus bundles with a representation
$\rho:\pi_1(\Sigma)\to SL(2,\mathbb{Z})$ inducing a
$Z[\pi_1(\Sigma)]$-module in terms of the $H_1(T^2)$ homology group
of the fiber.  Locally the target is a product of $M_9\times T^2$
but globally we cannot split the target from the base $\Sigma$ since
$T^2$ is the fiber of the non trivial symplectic torus bundle
$T^2\to\Sigma$. The formulation of the supermembrane in terms of
sections of the symplectic torus bundle with a monodromy  is a nice
geometrical structure to analyze global aspects of gauging
procedures on effective theories arising from M-theory. The allowed
classes of monodromy are those subgroups corresponding to the
elliptic, parabolic and hyperbolic inequivalent classes of $SL(2,\mathbb{Z})$
showed in the Section 2. But as already explained, the global
classification depends on the cohomology class of the fibration,  so
it is more refined at global level, i.e. there are more inequivalent
classes of symplectic torus bundles which may be related to different domain-wall solutions of supergravity.

\section{Classification of Symplectic Torus Bundles. }

Two conjugate representations $\rho$ and $U\rho U^{-1}$, with $U\in SL(2,\mathbb{Z})$, define $Z^2_{\rho}$ and $Z^2_{U\rho U^{-1}}$ modules with isomorphic cohomology groups $H^2 (\Sigma,Z^2_{\rho})\sim H^2 (\Sigma,Z^2_{U\rho U^{-1}})$. They define equivalent symplectic torus bundles. An equivalent way to see it is to consider the group of coinvariants associated to $\rho$ and $U\rho U^{-1}$. There is an isomorphism between the group of coinvariants associated to $\rho$ and to $U\rho U^{-1}$, they define equivalent symplectic torus bundles. In order to classify them, we must determine first the conjugacy classes of $SL(2,\mathbb{Z})$ and then then the associated coinvariants. Once this has been done the correspondence with the nine-dimensional gauged supergravities follows directly. $SL(2,\mathbb{Z})$ may be generated by $S$ and $ST^{-1}$ where
\bea
S=\begin{pmatrix} 0 & 1 \\ -1 & 0 \end{pmatrix}\quad\textrm{and}\quad T=\begin{pmatrix} 1 & 1 \\ 0 & 1 \end{pmatrix}.
\eea
Every conjugacy class of $SL(2,\mathbb{Z})$ can be represented by one of the following \cite{honda}
\bea
\begin{aligned}
& \pm S \quad \textrm{with Trace=0}.\label{r1}\\
& \pm T^{-1}S, \quad  \pm(T^{-1}S)^2,\quad \textrm{with} \quad\vert \textrm{Trace}\vert=1.\label{r2}\\
& \pm T^{n}, n\in \mathbb{Z} \quad \textrm{with} \quad\vert\textrm{Trace}\vert=2.\label{r3}\\
& \pm T^{r_0}S T^{r_1}S\dots T^{r_k}S \quad r_i\le -2, r_0<-2, i=1,\dots,k, \textrm{and}\quad \vert\textrm{Trace}\vert>2.\label{r4}
\end{aligned}
\eea
The representations:
\bea\label{con}
\begin{aligned}
&\rho(\alpha,\beta)= (\pm S)^{\alpha+\beta}\\
&\rho(\alpha,\beta)= (+ T^{-1}S)^{\alpha+\beta}\\
&\rho(\alpha,\beta)= (+ (T^{-1}S)^2)^{\alpha+\beta}\\
&\rho(\alpha,\beta)= (-\mathbb{I})^{\alpha+\beta}
\end{aligned}
\eea
 define finite subgroups isomorphic to $Z_4,Z_6,Z_3,Z_2$ respectively, associated to the monodromies
$\mathcal{M}_4,\mathcal{M}_6,\mathcal{M}_3,\mathcal{M}_2$ in \cite{dh2}. The representations
\bea
\rho(\alpha,\beta)=(-T^{-1}S)^{\alpha+\beta}\quad \textrm{and}\quad \rho(\alpha,\beta)=(-(T^{-1}S)^{2})^{\alpha+\beta}
\eea
define subgroups isomorphic to $Z_3$ and $Z_6$ respectively. The associated coinvariant groups are the trivial one and $Z_6$ respectively. In terms of the representation

\bea\label{hip}
\rho_{mn}(\alpha,\beta)=\begin{pmatrix}
-2mn+1 & 2mn^2+n\\ -m & 1+mn
\end{pmatrix}^{\alpha+\beta}
\eea
 with $m,n>0$ \cite{khan}, $[(T^{-1}S)^2]^{\alpha+\beta}$ is conjugate to $\rho_{31}(\alpha,\beta)$, $S^{\alpha+\beta}$ is conjugate to $\rho_{21}(\alpha,\beta)$ and $[T^{-1}S]^{\alpha+\beta}$ to $\rho_{11}(\alpha,\beta)$. The inequivalent symplectic torus bundles associated to $\rho_{mn}(\alpha,\beta)$ are $mn$ and all of them correspond to the torsion classes in $H^2 (B, Z_{\rho}^2)\equiv \mathbb{Z}_n\oplus \mathbb{Z}_m$ equivalently to the coinvariant group $Z_n\oplus Z_m$.
It is interesting that beyond the finite group cases $(\mathcal{M}_2,\mathcal{M}_3,\mathcal{M}_4,\mathcal{M}_6)$ associated to the elliptic case, there are monodromies defining non-finite subgroups associated to a finite number of symplectic torus bundle. For example $\rho_{41}(\alpha, \beta)$ is conjugate to $(-T^{-1})^{\alpha+\beta}\equiv \begin{pmatrix} -1 & 1\\ 0 & -1  \end{pmatrix}^{\alpha+\beta}$, which generates a non-finite subgroup, the associated number of symplectic torus bundles is finite, four in this case. The group of coinvariants is isomorphic to $Z_4$. For the parabolic conjugacy class $\vert \textrm{Trace}\vert =2$, there are two cases, the first one is associated to monodromies with a positive trace, they generate infinite symplectic torus bundles in correspondence to $\mathbb{Z}$, while the second case, with negative trace, generates a finite number of inequivalent  symplectic torus bundles. The group of coinvariants is always $Z_4$. In both cases the subgroups generated by the monodromy representation are not finite. If $mn>4$, $\textrm{Trace}\rho_{mn}(\alpha,\beta)<-2$. These are hyperbolic representations of $SL(2,\mathbb{Z})$. In this case there is a finite number of inequivalent symplectic torus bundles generated by non-finite subgroups.\newline

In this case, $mn>4$, the matrix $M\equiv \rho_{mn }(\alpha,\beta)$ (\ref{hip}) with $\alpha+\beta=1$ is conjugate, according to (\ref{con}) to $\pm T^{r_0}ST^{r_1}S\dots T^{r_k}S$, $r_i\le -2,r_0<2$, and $i=1,\dots,k$. In particular, we obtain for $n=1,m\ge 5$ that $M$ is conjugate to $-T^{-3}S(T^{-2}S)^{m-5}$. See appendix B. The group of coinvariants associated to the corresponding monodromy is $Z_m$, $m\ge 5$. There are $m$ inequivalent symplectic torus bundles corresponding to this monodromy. The sign is very relevant. For example, for $m=5$ $\rho_{51}(\alpha,\beta)= (-T^3S)^{\alpha+\beta}$ has a coinvariant group $Z_5$ while $(+T^3S)^{\alpha+\beta}$ has a trivial coinvariant group, with only the identity element. The latter case is not contained in (\ref{hip}), since it corresponds to positive trace.
\subsection{Gauge Fixing and Residual Symmetries.}
We may now consider the gauge freedom associated to the gauging of the abelian subgroups of $SL(2,Z)$. It corresponds to equivalent symplectic torus bundles arising in particular from conjugate representations $U\rho(\alpha,\beta)U^{-1}$, $U\in SL(2,Z)$. Two conjugate representations $\rho$ and $U\rho U^{-1}$, with $U\in SL(2,\mathbb{Z})$, define $Z^2_{\rho}$ and $Z^2_{U\rho U^{-1}}$ modules with isomorphic cohomology groups $H^2 (\Sigma,Z^2_{\rho})\sim H^2 (\Sigma,Z^2_{U\rho U^{-1}})$. They define equivalent symplectic torus bundles. An equivalent way to see it is to consider the group of coinvariants associated to $\rho$ and $U\rho U^{-1}$.  In fact, the group $H^2 (\Sigma, Z^2_{\rho})$ is isomorphic, via Poincare duality, to the coinvariants group associated to $\rho$. There is then an isomorphism between the group of coinvariants associated to $\rho$ and to $U\rho U^{-1}$, they define equivalent symplectic torus bundles.  Given $\mathcal{Q} \equiv \begin{pmatrix} p\\q \end{pmatrix}\in H_1 (T^2)$, the group of coinvariants of monodromy $\rho$ is the abelian group of equivalence classes
\bea
\{\mathcal{Q}-\Lambda\mathcal{\widehat{Q}}-\mathcal{\widehat{Q}}\}
\eea
for any $\Lambda\in \rho$ and any $\mathcal{\widehat{Q}}=\begin{pmatrix}\widehat{p}\\ \widehat{q}\end{pmatrix}\in H_1 (T^2)$. It follows that this class is mapped to the class associated to $U\mathcal{Q}$ under the representation $U\rho U^{-1}$:
\bea
\{U\mathcal{Q}-U\Lambda U^{-1}\mathcal{\widetilde{Q}}-\mathcal{\widetilde{Q}}\}
\eea
where $\mathcal{\widetilde{Q}}=U\mathcal{\widehat{Q}}$, but any $\mathcal{\widetilde{Q}}\in H_1(T^2)$ may always be expressed as $U\mathcal{\widehat{Q}}$ for some other $\mathcal{\widehat{Q}}\in H_1(T^2)$, since $U$ is invertible. There is then an isomorphism between the group of coinvariants associated to $\rho$ and to $U\rho U^{-1}$, they define equivalent symplectic torus bundles.
\newline
We may choose $U$ in order to leave freezed the winding matrix under the action of the monodromy transformation. The gauge fixing procedure goes as follows. We re-arrange the winding matrix as $M=\begin{pmatrix}m_1 & l_1 \\ m_2 & l_2&\end{pmatrix}$, with $det M=n$. Under the symmetry of Section 3 it transforms as
\begin{equation}
\begin{pmatrix}
s_1 & s_2\\ s_3 & s_4
\end{pmatrix}\begin{pmatrix}
m_1 & l_1\\ m_2 &l_2
\end{pmatrix}\Lambda^{-1}
\end{equation}
The $SL(2,Z)$ symmetry associated to the base manifold may be interpreted as having independence on the basis of homology on the base manifold. In fact, the winding matrix is associated to a particular basis of homology. Hence, since the change of homology basis corresponds to a $SL(2,\mathbb{Z})$ transformations, the theory should only depend on the equivalence classes constructed from the application from the left by a $SL(2,\mathbb{Z})$ matrix:
\begin{equation}
\begin{pmatrix}
s_1 & s_2\\ s_3 & s_4
\end{pmatrix}\begin{pmatrix}
m_1 & l_1\\ m_2 &l_2
\end{pmatrix}.
\end{equation}
Under this transformation the winding matrix may always be reduced to the canonical form
\begin{equation}
\begin{pmatrix}
\lambda_1 & 0\\ \beta & \lambda_2
\end{pmatrix}
\end{equation}
with $\lambda_1\lambda_2=n$ the central charge defined in Section 2, and $\vert \beta\vert \le \lambda_1/2$. In particular, if $\lambda_1=n$, $\lambda_2=1$ then $|\beta| \le \frac{n}{2}$. we notice that in addition to the central charge integer $n$ there are additional degrees of freedom represented by the integer $\beta$. We may now consider the supermembrane formulated as a symplectic torus bundle with monodromy $U\rho(\alpha,\beta)U^{-1}$. The action on the winding matrix is given by
\begin{equation}
\begin{pmatrix}
\lambda_1 & 0\\ \beta &\lambda_2
\end{pmatrix} U\rho^{-1}U^{-1}.
\end{equation}

We may also act from the left by a $SL(2,\mathbb{Z})$ matrix which we take of the form $V^{-1}\rho^{*}V$. We can take $U$ and $V$ both $SL(2,\mathbb{Z})$ matrix in order to rewrite the winding matrix in  form which is left invariant under the action of $\rho^{*}$ and $\rho^{-1}$.  For example if we take the monodromy
\bea
\rho(\alpha,\beta)= \begin{pmatrix} a & nb_1\\ c &d\end{pmatrix}^{\alpha+\beta}\in SL(2,\mathbb{Z})
\eea
associated to a supermembrane with central charge $n$, for particular values of $a,b,c,d$ and $n$, this includes elliptic, parabolic and hyperbolic monodromies. Then we can take
\bea
\rho(\alpha,\beta)^{*}= \begin{pmatrix} a & b_1\\ nc &d\end{pmatrix}
\eea
and $V,U$ such that
\bea
V\begin{pmatrix} \lambda_1 & 0\\ \beta &\lambda_2\end{pmatrix}U=\begin{pmatrix} 1& 0\\ 0 &n\end{pmatrix}
\eea
Then
\bea
\rho^{*}(\alpha,\beta)\begin{pmatrix}1& 0\\ 0 & n\end{pmatrix}\rho^{-1}(\alpha,\beta)=\begin{pmatrix}1& 0\\ 0 & n\end{pmatrix}
\eea
We then have
\begin{equation}
V^{-1}\rho^{*}V\begin{pmatrix}
\lambda_1 & 0\\ \beta &\lambda_2
\end{pmatrix} U\rho^{-1}U^{-1}=\begin{pmatrix}
\lambda_1 & 0\\ \beta &\lambda_2
\end{pmatrix}
\end{equation}
that is, the winding matrix is left invariant under the monodromy $\rho(\alpha,\beta)$ provided we consider an associated abelian representation of $SL(2,\mathbb{Z})$ acting on the homology of the base manifold.
Having established the gauge fixing procedure arising from conjugate representations $U\rho(\alpha,\beta)U^{-1}$, we may now ask what is the residual symmetry of the supermembrane on that symplectic torus bundle with monodromy $U\rho (\alpha,\beta)U^{-1}$. The residual symmetry must leave invariant the elements of the coinvariant group of the monodromy. It must act as the identity on the coinvariant group. Consequently it is the same abelian group defining the monodromy. In distinction, a group that commutes with the monodromy group maps the coinvariant group into itself, but it does not need to act as the identity. The latest corresponds to the residual symmetry of a theory when on considers the collection of bundles associated to a given monodromy.  The collection procedure occurs when we construct gauged supergravities in 9D from the 11D compactified supermembrane theory on the symplectic torus bundle with the central charge condition.
\newline
We may finally express the hamiltonian of the supermembrane with central charges on a symplectic torus bundle with monodromy $\rho(\alpha,\beta)$ in the following way,
\bea\label{z}
 \begin{aligned}
H&=\int_{\Sigma}T^{2/3}\sqrt{W}[\frac{1}{2}(\frac{P_{m}}{\sqrt{W}})^{2}+\frac{1}{2}
(\frac{P\overline{P}}{W})+
\frac{T^2}{4}\{X^{m},X^{n}\}^{2}+\frac{T^2}{2}\mathcal{D}X^{m}\overline{\mathcal{D}}X^{m}+\frac{T^2}{8}\mathcal{F}\overline{\mathcal{F}}
\\ & \nonumber -\int_{\Sigma}T^{2/3}
\sqrt{W} [\overline{\Psi}\Gamma_{-}
\Gamma_{m}\{X^{m},\Psi\}+1/2 \overline{\Psi}\Gamma_{-} \overline{\Gamma}
\{X,\Psi\}+1/2 \overline{\Psi}\Gamma_{-} \Gamma
\{\overline{X},\Psi\},
 \end{aligned}
 \eea
 subject to the first class constraints. We denote 
 \begin{equation}
 \begin{aligned}
 &\mathcal{D}\circ= D\circ +\{A,\circ\},\\
 &\mathcal{F}= D\overline{A}-\overline{D}A+\{A,\overline{A}\}
 \end{aligned}
 \end{equation}
where  
\bea D=\frac{\epsilon^{ab}}{\sqrt{W}} 2\pi R(l_r+m_r\tau)\theta_{r}^{s}
\partial_{a}\widehat{X}^{s}\partial_{b}\eea
and the matrix $\theta$ is given by 
\bea
\theta=(V^{-1}(\rho^*)^{-1}V)^{T}.
\eea
The matrix $\theta$ was derived by the sculpting approach. We have obtained its explicit expression here from the gauge fixing procedure introduced in this section. As mentioned before in this section $D$ and $A$ acquire a phase factor $e^{i\varphi_{\rho}}$ as a consequence of the monodromy.  The hamiltonian is manifestly invariant under this transformation. The moduli $R$ and $\tau$ transform as (\ref{17}) where $\Lambda= U\rho U^{-1}$. The factor $\theta_s^r$ in the expression of $D$ arises from the transformation of the basis of harmonic one-forms. It can be also interpreted as a transfromation of the winding matrix with components $l_r$ and $m_r$, $r=1,2$. If we take this point of view the winding numbers belong then to an element of the coinvariant group associated to the monodromy $V^{-1}(\rho^{*})V$ acting on the cohomology of the base manifold while the KK charges belong to an element of the coinvariant group of monodromy $\rho$. The mass squared formula remains then invariant under transitions on the symplectic torus bundle provided we interpret the winding numbers and KK charges as equivalence classes of the corresponding coinvariant groups.

\section{Gauging of the Trombone Symmetry on the Supermembrane.}

In the previous section we showed that the supermembrane with
central charges may be formulated on a symplectic torus bundle with
a nontrivial $SL(2,Z)$ monodromy. Corresponding to each monodromy we
obtain the gauging of an abelian subgroup of $SL(2,\mathbb{Z})$, the
isotopy group of symplectomorphisms preserving the symplectic 2-form
introduced in the construction of the supermembrane theory with
central charges. The monodromy defined as the homomorphism from
$\Pi_1(\Sigma)\to \Pi_0(G)\approx SL(2,\mathbb{Z})$ was constructed
in terms of parabolic, elliptic and hyperbolic $SL(2,\mathbb{Z})$
matrices. We are going to show that there is also a supermembrane
theory with central charges formulated on a symplectic torus bundle
with a monodromy corresponding to the gauging of the trombone
symmetry introduced in the context of supergravity
\cite{cremmer}. See Section 2.  The first step will be to consider the
supermembrane formulated on a symplectic torus bundle with trivial
monodromy and obtain the transformation law of
the mass squared formula presented in section 3 under the scaling
symmetry. We first follow the approach \cite{cremmer} and work out
the general compensator in the context of the supermembrane theory.
The second step will be to gauge the trombone symmetry in M-theory.

\subsection{The Trombone Symmetry on the Compactified M2 with Central Charges}

Let us obtain the transformation law of the mass squared formula
presented in section 3 under the scaling symmetry. Following the
lines of \cite{cremmer}, we are going to generalize the compensating
transformation for arbitrary values of the moduli $\tau$.

\paragraph {The General Form of the Compensating Transformation:} We consider a integer lattice of KK charges parametrized by
$
Q=\begin{pmatrix}
p\\q
\end{pmatrix}
$.
The geometrical interpretation of $Q$ is in terms of the elements
of the homology group $H_1 (T^2)$ of the fiber, which is a 2-torus.
Under the U-duality transformation (\ref{17}) $Q\to\Lambda Q$ with
$\Lambda\in SL(2,\mathbb{Z})$ with the corresponding transformation
of the moduli parameters as stated in section 3. We are interested
in the most general transformation mapping $Q_i\to Q_j:
Q_j=\Lambda_{ij}Q_i$. For a given $Q_i$ we define $\Lambda_i\in
SL(2,\mathbb{Z}): \Lambda_iQ_0=Q_i$ where $ Q_0=\begin{pmatrix} 1\\0
\end{pmatrix}
$. $\Lambda_i$ is not unique, its most general expression is
$\Lambda_i g$ where $g=\begin{pmatrix} 1 & m\\ 0 & 1\end{pmatrix}$
for any integer $m\ne 0$, and $g\in H$ is the Borel group of
parabolic $SL(2,\mathbb{Z})$ matrices.  We then have
$\Lambda_{ji}=\Lambda_jg\Lambda_i^{-1}\quad \textrm{for any } g\in
H.$ Under composition we have \bea
\Lambda_{kj}\Lambda_{ji}=\Lambda_{ki}. \eea

For $\Lambda_{ji}\in SL(2,\mathbb{Z})$ acting on $Q_i$ there is an
associated transformation of the moduli parameters as stated in
section 3. The mass formula is invariant under the overall
transformation. We consider equivalence classes of matrices
$\Lambda_{ji}$: two elements of the class differ in an element $g\in
H$. We denote the class $\widetilde{\Lambda}_{ij}$. We may now
introduce the compensator in the approach of \cite{cremmer}. The
following result is valid: for each equivalence class
$\widetilde{\Lambda}_{ji}$ there exists a unique matrix $H_{ji}\in
GL(2,\mathbb{R})$, $H_{ji}=M_{ji}\Lambda_{ji}$ and a unique complex
number $h_{ji}\in\mathbb{C}$ such that
\begin {itemize}
\item {(i)} $H_{ji} Q_i= Q_j$\\
\item{(ii)} $H_{ji} \begin{pmatrix}
\tau&\\1\end{pmatrix}=h_{ji}\begin{pmatrix}
\tau&\\1\end{pmatrix}$
\end{itemize}
$H_{ji}$ and $h_{ji}$ depend only on the equivalence class, it is independent of
$g\in H$. In distinction, the compensator $M_{ji}$ depends
explicitly on $g\in H$. Relation ii) is equivalent to the following
sequence of transformations:
\begin{equation}
\tau\stackrel{\Lambda_{ji}}{\rightarrow}\tilde{\tau}\stackrel{M_{ji}}{\rightarrow}\tau
\end{equation}
where $\tau\to\widetilde{\tau}$ is the Moebius transformations associated to $\Lambda_{ji}\in SL(2,\mathbb{Z})$. The general expression of the $H_{ji}$ matrix is,
\begin{equation}
\begin{aligned}
H_{ji}=\begin{pmatrix}
-\frac{p_j}{q_j}u+\frac{q_i}{q_j}\mathcal{C} &\quad \frac{p_j}{q_i}+\frac{p_ip_j}{q_iq_j}u-\frac{p_i}{q_j}\mathcal{C}\\  -u & \frac{q_j}{q_i}+\frac{p_i}{q_i}u \end{pmatrix}
\end{aligned}
\end{equation}
with $u=\frac{(p_jq_i-p_iq_j)}{\vert p_i-q_i\tau\vert^2}$,
$\mathcal{C}=det M_{ji}=\frac{\vert p_j-q_j\tau \vert^2}{\vert
p_i-q_i\tau \vert^2}$ and
$h_{ji}=\frac{p_j-q_j\overline{\tau}}{p_i-q_i\overline{\tau}}$,
where $\overline{\tau}$ is the complex conjugate of $\tau$. It then
follows that the compensator $M_{ji}$ depends explicitly on $g\in H$
since $M_{ji}=H_{ji}\Lambda_{ji}^{-1}$. Although $H_{ji}\in
GL(2,\mathbb{R})$, the non-linear transformation maps integer
charges $Q_i$ into integer charges $Q_j$, as it should in order to
satisfy the charge quantization condition. It is straightforward to
show that $H_{ji}$ defines a non-linear realization of the
$SL(2,\mathbb{Z})$ group. In fact, if \bea \widetilde{\Lambda}_{21}\to
H_{21},\quad \widetilde{\Lambda}_{32}\to H_{32},\quad \widetilde{\Lambda}_{31}\to H_{31}
\eea then $ H_{21}Q_1=Q_2,\quad H_{32}Q_2=Q_3 $ hence
$H_{32}H_{21}Q_1=Q_3$. Analogously, \bea
H_{32}H_{21}\begin{pmatrix}\tau\\ 1\end{pmatrix}=
\lambda_{32}\lambda_{21}\begin{pmatrix}\tau\\
1\end{pmatrix}=\lambda_{31}\begin{pmatrix}\tau\\ 1\end{pmatrix} \eea

The uniqueness of the transformation then implies $H_{31}=
H_{32}H_{21}$.\newline $H_{ji}$ realizes then a nonlinear representation of
$SL(2,\mathbb{Z})$ and it represents the trombone symmetry at the
quantum level.

\paragraph{The Mass Operator Transformation under Trombone Symmetry.}
 Having determined the transformation law for the KK charges and the complex moduli $\tau$ we may now consider the transformation of the other moduli R, and the winding matrix. From (\ref{17}) we know their transformation law under $\Lambda_{ji}\in SL(2,\mathbb{Z})$, we may now determine the compensator action on them . We will do so by imposing the condition that the hamiltonian remains invariant under its action. The transformation for the complex moduli $\tau$ may be  re-written as:
\bea
\begin{pmatrix}\label{tau}
\tau\\1\end{pmatrix}\stackrel{\frac{\Lambda_{ji}}{l_{ji}}}{\rightarrow}\begin{pmatrix}
\tau^{'}\\1\end{pmatrix}\stackrel{\frac{l_{ji}}{h_{ji}}M_{ji}}{\rightarrow}\begin{pmatrix}
\tau\\1\end{pmatrix}\eea

where $l_{ji}\equiv c\tau+d$ and $\Lambda\in SL(2,\mathbb{Z})$, see (\ref{17})
while $\frac{1}{\vert h_{ji}\vert}M_{ji}\in SL(2,\mathbb{R})$ and
$h_{ji}$ was defined as in the previous section. The harmonic sector of
the supermembrane may be expressed as \bea 2\pi R
(d\widehat{X}^1,d\widehat{X}^2)\begin{pmatrix} m_1 & l_1\\ m_2 &
l_2\end{pmatrix}\begin{pmatrix} \tau\\1\end{pmatrix}. \eea Under the
first transformation in the composition \ref{tau} the factor $\vert
l_{ji}\vert^{-1}$ is canceled by the transformation of $R$: \bea R
\stackrel{\vert l_{ji}\vert}{\rightarrow} R^{'}= R\vert
l_{ji}\vert. \eea We must then consider \bea R^{''}=\frac{R^{'}}{\vert
l_{ji}\vert} \eea in order to compensate the factor $\vert
l_{ji}\vert$ in the second transformation in \ref{tau}. We then have
\bea R \to R^{'}\to R. \eea Finally, under $\Lambda_{ji}$ the
winding matrix transform as: \bea
\begin{pmatrix}
m_1 & l_1 \\m_2 & l_2\end{pmatrix}\to
\begin{pmatrix}
m_1^{'} & l_1^{'} \\m_2^{'} & l_2^{'}\end{pmatrix}=\begin{pmatrix}
m_1 & l_1 \\m_2 & l_2\end{pmatrix}\Lambda_{ji}^{-1}\eea
Consequently, the compensating action must be

\bea
\begin{pmatrix}
m_1^{'} & l_1^{'} \\m_2^{'} & l_2^{'}\end{pmatrix}\stackrel{\Lambda_{ji}^{-1}}{\rightarrow}
\begin{pmatrix}
m_1 & l_1 \\m_2 & l_2\end{pmatrix}\eea
in order to have
an invariant hamiltonian under that action. We notice that the
harmonic sector is not invariant but its contribution together with
the one of its complex conjugate yields an invariant hamiltonian.
The winding term in the mass formula also remain invariant while the
KK therm varies according to:

 \bea
 \frac{\vert p_i-q_i\tau \vert}{R Im{\tau}}\to \frac{\vert p_j-q_j\tau \vert}{R Im{\tau}}
 \eea

\subsection{Gauging the Trombone}

We may finally consider the gauging of the trombone symmetry. The
main point in the construction is the geometrical description of the
KK charges $(p,q)$ in terms of the elements of the homology group
$H_1(T^2)$ of the fiber $T^2$. The homomorphism $\Pi_1(\Sigma)\to
\Pi_0(G)\approx SL(2,\mathbb{Z})$ determines a representation $\rho:
\Pi_1(\Sigma)\to SL(2,\mathbb{Z})$. If we denote
$\rho(\alpha,\beta)\in SL(2,\mathbb{Z})$ the element of
$SL(2,\mathbb{Z})$ associated to $(\alpha,\beta)\in\pi_1(\Sigma)$, its
action on $H_1(T^2)$ yields \bea Q_j=\rho(\alpha,\beta)Q_i \eea

Form section (6.1) we then conclude that
$\rho(\alpha,\beta)=\Lambda_{ji}$ and there exists an associated
non-linear representation realized in terms of the matrix $H_{ji}$.
The monodromy is then constructed with this non-linear
representation of $SL(2,\mathbb{Z})$. we notice that the
$Z[\Pi_1(\sigma)]$-module is the same as the one arising from the
linear representation  $\rho$, however its action on $\tau,R$ and
the winding matrix is different since their transformation is done
in terms of $H_{ji}$ matrices. We thus obtain a different global
structure for the supermembrane on this symplectic torus bundle.
Following the analysis of section 5, the hamiltonian of the
supermembrane is well-defined on this symplectic torus bundle. We
notice that the $(p,q)$ charges in the KK term of the mass squared
formula do not have arbitrary values.  In fact the only allowed
values are the ones determined from the $Z_{\rho}^2$-module. In
order to obtain the invariance of the mass squared formula we may
consider summation on all the $(p,q)$ values allowed by the
$Z_{\rho}^2$- module. One arrives to the family of symplectic torus
bundle whose monodromy realizes the gauging of the trombone
symmetry.



\section{T-duality in the Supermembrane Theory}
In this section we introduce the T-duality transformations for the supermembrane theory. This goes beyond the T-duality of superstring theory. In fact, the latter may be directly obtained from the membrane theory by freezing membrane degrees of freedom and quantizing the remaining string states \cite{sl2z}. In this section we present the T-duality of the full degrees of freedom of the supermembrane, when formulated on a dual symplectic torus bundle (i.e. a symplectic torus bundle defined under the T-duality transformation acting on the moduli). It acts on the moduli as well as on the bosonic and fermionic fields. We will see that T-duality become a natural symmetry of the theory that fixes the scale of energy of the supermembrane tension $T$. The T-duality transformation is a nonlinear map which interchange the winding modes $\mathbb{W}$, previously defined associated to the cohomology of the base manifold with the KK charges, $Q =(p,q)$ associated to the homology of the target torus together with a transformation of the real moduli $R \to \frac{1}{R}$ and complex moduli $\tau \to \widetilde{\tau}$, both in a nontrivial way. In the following all transformed quantities under T-duality are denoted by a tilde, to differenciate from other symmetries.
Given a winding matrix $\mathbb{W}$ and KK modes there always exists an equivalent winding matrix $\mathbb{W}^{'}=\begin{pmatrix}l_1^{'} & l_2^{'}\\ m_1^{'} & m_2^{'} \end{pmatrix}$,  under the $SL(2,\mathbb{Z})$ symmetry (\ref{16}) such that for KK charges  $Q=\begin{pmatrix}p\\ q \end{pmatrix}$,
\bea\label{ppp}
\begin{pmatrix}l_1^{'}\\ m_1^{'} \end{pmatrix}= \Lambda_0 \begin{pmatrix}p\\ q\end{pmatrix}
\eea
where $\Lambda_0=\begin{pmatrix}\alpha & \beta\\ \gamma & \delta \end{pmatrix} \in SL(2,\mathbb{Z})$ with $\alpha=\delta$. This is an intrinsic relation between the equivalence classes of winding matrices and KK modes. In fact, it is preserved under a U-duality transformation (\ref{16}):

\begin{equation}
\begin{aligned}
\begin{pmatrix}l_1^{'}\\ m_1^{'} \end{pmatrix}\longrightarrow&\begin{pmatrix}\widehat{l}_1\\ \widehat{m}_1\end{pmatrix}= \begin{pmatrix}a& -b\\ -c& d \end{pmatrix}\begin{pmatrix}l_1^{'}\\ m_1^{'} \end{pmatrix}\\
\begin{pmatrix}p\\ q \end{pmatrix}\longrightarrow &\begin{pmatrix}\widehat{p}\\ \widehat{q}\end{pmatrix}= \begin{pmatrix}a& b\\ c& d \end{pmatrix}\begin{pmatrix}p\\ q \end{pmatrix}\\
\end{aligned}
\end{equation}
Hence
\begin{equation}
\begin{pmatrix}
\widehat{l}_1\\ \widehat{m}_1\end{pmatrix}=\mathbb{M}\begin{pmatrix}\widehat{p}\\ \widehat{q}\end{pmatrix}
\end{equation}
where
\begin{equation}
\mathbb{M}=\begin{pmatrix}a& -b\\ -c& d \end{pmatrix}\Lambda_0\begin{pmatrix} a& b\\ c& d
\end{pmatrix}^{-1}.
\end{equation}
The matrix $\mathbb{M}\in SL(2,\mathbb{Z})$ and has equal diagonal terms, provided $\Lambda_0$ has $\alpha=\delta$. In order to define the T-duality transformation we introduce the following \cite{sl2z}(47)  dimensionless variables
\bea
\mathcal{Z}:= TA\widetilde{Y}\quad \mathcal{\widetilde{Z}}:= T \widetilde{A}Y
\eea

where $T$ is the supermembrane tension, $A = (2\pi R)^2 Im\tau$ is the area of the target torus and $Y=\frac{R Im\tau}{\vert q\tau -p\vert}$. The tilde variables $\widetilde{A}, \widetilde{Y}$ are the transformed quantities under the T-duality\footnote{This definition can be more naturally understood in terms of a vector of the 2-torus moduli $\mathcal{V}=(TA,\mathcal{Z}/Y)$ defined in terms of the moduli $(R,\tau)$ as
\bea\label{za}
\widetilde{\mathcal{V}}=\Omega \mathcal{V}
\eea
being  $\Omega=\begin{pmatrix}0 & 1 \\ 1 & 0\end{pmatrix}$.}. See (\ref{trans}) for the explicit value of $\mathcal{Z}$. The T-duality transformation we introduce is given by:
\bea\label{tt1}
\begin{aligned}
\textrm{The moduli}:& \quad \mathcal{Z}\widetilde{\mathcal{Z}}=1,\quad\widetilde{\tau}=\frac{\alpha \tau+\beta}{\gamma\tau +\alpha}; \\
\textrm{The charges}:& \begin{pmatrix}\widetilde{p}\\ \widetilde{q}\end{pmatrix}=\Lambda_0 \begin{pmatrix} p\\ q\end{pmatrix},
\begin{pmatrix}\widetilde{l}_1& \widetilde{l}_2\\ \widetilde{m}_1 & \widetilde{m}_2\end{pmatrix}=\Lambda_0^{-1}\begin{pmatrix}l_1^{'} & l_2^{'}\\ m_1^{'} & m_2^{'} \end{pmatrix}.
\end{aligned}
\eea
We notice that the T-duality transformations for the winding matrix, having $\Lambda_0$ equal diagonal terms, becomes of the same form as in (3.22). The main difference is that $\Lambda_0$ is determined in terms of the winding and KK modes, defining a nonlinear transformation on the charges of the supermembrane, while (3.22) is a linear transformation on them. With the above definition of T-duality transformation we have

\begin{equation}\label{a3}
\begin{aligned}
&\begin{pmatrix}
p\\ q\end{pmatrix}\to \begin{pmatrix}
\widetilde{p}\\ \widetilde{q}\end{pmatrix}=\begin{pmatrix}
l_1^{'}\\ m_1^{'}\end{pmatrix}\\
&\begin{pmatrix}
l_1^{'}\\ m_1^{'}\end{pmatrix}\to\begin{pmatrix}
\widetilde{l}_1^{'}\\
\widetilde{ m}_1^{'}\end{pmatrix}=\begin{pmatrix} p\\ q \end{pmatrix}
\end{aligned}
\end{equation}
See the Appendix A for the construction of $\Lambda_0$.
That is, the KK modes are mapped onto the winding modes and viceversa. The property together with the condition $Z\widetilde{Z}=1$ ensure that $(\textrm{T-duality})^2= \mathbb{I}$, the main property of T-duality.  The explicit transformations of the real modulus, obtained from the above T-duality transformation is
\begin{equation}\label{trans}
\widetilde{R}= \frac{\vert \gamma\tau+\alpha\vert \vert q\tau-p\vert ^{2/3}}{T^{2/3}(Im \tau)^{4/3}(2\pi)^{4/3}R},\qquad \\ \textrm{with}\quad\widetilde{\tau}=\frac{\alpha \tau+\beta}{\gamma\tau +\alpha} \quad \textrm{and}\quad \mathcal{Z}^2=\frac{T R^3 (Im\tau)^2}{\vert q\tau-p\vert}
\end{equation}
The winding modes and KK charge contribution in the mass squared formula transform in the following way:
\bea
\begin{aligned}
Tn^2 A^2 &= \frac{n^2}{\widetilde{Y}^2}\mathcal{Z}^2\\
\frac{m^2}{Y^2}&= T^2m^2 \widetilde{A}^2\mathcal{Z}^2
\end{aligned}
\eea
To see how the $H_1$ (\ref{28}) transforms under T-duality it is important to realize the transformation rules for the fields,
 \bea
 \begin{aligned}
 dX^m&=u d\widetilde{X}^m,\quad
 d\widetilde{X}=u e^{i\varphi}dX,\quad
 A=u e^{i\varphi}\widetilde{A}\\ \textrm{and}\qquad
 \Psi&=u^{3/2} \widetilde{\Psi},\quad
 \overline\Psi= u^{3/2}\widetilde{\overline\Psi}
 \end{aligned}
 \eea
 Where $u=\mathcal{Z}^2=\frac{R\vert \gamma\tau+\alpha\vert}{\widetilde{R}}$, $\varphi$ was defined in (\ref{17}) and $dX=dX^1+idX^2$ and respectively, its dual $d\widetilde{X}$ is\bea
\begin{aligned}
d\widetilde{X}=2\pi \widetilde{R}[(\widetilde{m}_1\widetilde{\tau}+\widetilde{l_1})d\widehat{X}^1+(\widetilde{m}_2\widetilde{\tau}+\widetilde{l}_2)d\widehat{X}^2]
\end{aligned}
\eea
The phase $e^{i\varphi}$ cancels with the h.c. the transformation of the Hamiltonian.
 The relation between the hamiltonians through a T-dual transformation is
 \bea
 H=\frac{1}{\widetilde{\mathcal{Z}}^{8}}\widetilde{H},\quad \widetilde{H}= \frac{1}{\mathcal{Z}^8}H.
 \eea

 We thus obtain for the mass squared formula the following identity,
 \bea
 M^2 = T^2 n^2 A^2 + \frac{m^2}{Y^2}+ T^{2/3}H =\frac{1}{\widetilde{\mathcal{Z}}^2}(\frac{n^2}{\widetilde{Y}^2}+ T^2 m^2 \widetilde{A}^2)+ \frac{T^{2/3}}{\widetilde {\mathcal{Z}}^8}\widetilde{H}.
\eea
%

\subsection{T-Duality on Symplectic Bundles}
 There is bijective relation between the symplectic torus bundles with monodromy $\rho(\alpha,\beta)$ and the elements of the cohomology group $H_2 (\Sigma, Z_{\rho})$ of the base manifold $\Sigma$ with coefficients on the module $Z_{\rho}^{2}$, and hence with the elements of the coinvariant group associated to the monodromy group $G$. That is each equivalence class
 \bea
 \{\mathcal{Q}+g\widehat{\mathcal{Q}}-\widehat{\mathcal{Q}}\},
 \eea
for any $g\in G$ and $\widehat{\mathcal{Q}}\in H_1 (T^2)$, characterizes one symplectic torus bundle. In the formulation of the supermembrane on that geometrical structure $\mathcal{Q}$ are identified with the KK charges. The action of $G$, the monodromy group, leaves the equivalence class invariant. $G$ acts as the identity on the coinvariant group. We now consider the duality transformation introduced previously. It interchanges KK modes $\mathcal{Q}$ into components of the winding matrix through the relation (\ref{ppp})
\bea
\begin{pmatrix}l_1\\ m_1\end{pmatrix}= \Lambda_0 \begin{pmatrix}p\\ q\end{pmatrix}
\eea
Under the duality transformation the equivalence class transform as

\bea
 \{\Lambda_0\mathcal{Q}+(\Lambda_0 g \Lambda_0^{-1})\Lambda_0\widehat{\mathcal{Q}}-\Lambda_0\widehat{\mathcal{Q}}\},
 \eea
 hence for the dual bundle it holds,
 \bea
 \{\Lambda_0\begin{pmatrix}l_1\\ m_1\end{pmatrix}+(\Lambda_0 g \Lambda_0^{-1})\begin{pmatrix}\widehat{l_1}\\ \widehat{m_1}\end{pmatrix}-\begin{pmatrix}\widehat{l_1}\\ \widehat{m_1}\end{pmatrix}\},\eea
 That is, as an element of the coinvariant group of $\Lambda_0 G \Lambda_0^{-1}$. We then conclude that the duality transformation, in addition to the transformation on the moduli $R,\tau$, also maps the geometrical structure onto an equivalent symplectic torus bundle with monodromy $\Lambda_0 G\Lambda_0^{-1}$. We notice that the transformation depends crucially on the original equivalence class of the coinvariant group. So for a nonequivalent symplectic torus bundle the dual transformations is realized with a different $SL(2,Z)$ matrix $\Lambda_0$. Consequently, this dual transformation between supermembrane on symplectic torus bundles cannot be seen at the level of supergravity theory which only distinguish the monodromy group but not its coinvariant structure.
\newline
Now we are in position to determine the T-duality as a natural symmetry for the family of
supermembranes with central charges. We take: \bea\label{la} \widetilde{Z}=Z=1\Rightarrow
T_0=\frac{\vert q\tau-p\vert}{R^3 (Im\tau)^2}. \eea
It imposes a relation between the energy scale of the tension of the supermembrane and the moduli of
the torus fiber and that of its dual. Indeed we can think in two different ways: given the values of the moduli it fixes the allowed tension $T_0$ or on the other way around, for a fixed tension $T_0$, the radius, the Teichmuller parameter of the 2-torus, and the KK charges satisfy (\ref{la}).
When this T-duality extended to M-theory acts on the stringy states
of the supermembrane with central charges wrapping on a $\widetilde{T}^{2}$ one recovers the
standard T-duality relations in string theory \cite{sl2z}. The contribution of the stringy states of the supermembrane with central charges
wrapping on a dual $\widetilde{T}^{2}$ torus was already found in
\cite{sl2z}.
At the level of supergravity the structure of the fiber bundle base
manifold of the supermembrane with central charges is lost and a remanent of it appears as nonvanishing components
of the 3-form, which  for the supermembrane in the LCG corresponds
to $C_{-rs}$ \cite{dwpp-background}. Following the lines of the noncommutative torus of \cite{connes}\footnote{The work of \cite {connes} is mainly done in flat space with Moyal star product in which the noncommutative parameter is given by the 2-form, however as it is signalled in the paper, it can be  generalized to curved manifolds, for which the star product is changed to a deformation quantization star product ( for example in our case it corresponds to a Fedosov-like product) and then, an additional choice of Poisson structure appears.}, we can interpret $C_{-rs}=F_{rs}$ in our case the nondegenerate 2-form associated to the central charge condition, then $\int_{\Sigma} F_{rs}=n$ and at the level the noncommutative structure of the 2-torus in string theory the nonvanishing three form corresponds to the
presence of nonvanishing $B_{ij}$ field \cite{lust} in the closed string sector. The formulation of the supermembrane in the presence of nonvanishing 3-form has been analyzed in \cite{dwpp-background}. In our formulation there is a particularity, since the magnetic field on the worldvolume of the supermembrane induced by the monopole contribution is nonconstant and consequently it should be associated to a
nonvanishing 4-form flux $G=dC$ in 11D.  In \cite{lust} the double T-duality is realized for
the the closed strings sector and its associated noncommutativity,
it would be interesting to see if there is a connection with our
results.

\section{Discussion and Conclusions}
We showed that the formulation of the supermembrane in terms of
sections of the symplectic torus bundle with a monodromy  is a
natural way to understand the M-theory origin of the gauging
procedures in supergravity theories \cite{gmmpr}. Its low energy
limit corresponds to the type II $SL(2,\mathbb{R})$ gauged
supergravities in 9D. We have explicitly shown the relation with the type
IIB gauged sugras in 9D. The global description is a realization of the
sculpting mechanism found  in \cite{sculpting} and it is associated to
the inequivalent classes of symplectic torus bundles with
monodromies in $SL(2,\mathbb{Z})$. The geometrical description of these kind of bundles has been developed in \cite{khan}. As already
conjectured in \cite{sculpting} we claim that the following diagramme
applies:
\begin{equation}
\begin{diagram}
\node{\textrm{Compactified M2($n=0$) in $X_9\times T^2$} } \arrow[2]{e,t}{\textit{'Sculpting'}}
     \arrow{s,l}{\textrm{Low Energies}}
        \node[2]{\textrm{M2 with central charges ($n\ne 0$) in $X_9\times T^2$}} \arrow{s,r}{\textrm{Low Energies}}\\
\node{\textrm{Type II Maximal Supergrav. 9D}}\arrow[2]{e,t}{\textit{Noether}}
   \node[2]{\textrm{Type II Gauged Supergravities 9D}}
        \label{eq:diag2}
\end{diagram}
\end{equation}

The supermembrane without any extra topological condition compactified on a 2-torus is a gauge theory on a trivial principle bundle with structure group of the symplectic group homotopic to the identity. The supermembrane with nontrivial central charge is also invariant under the isotopy group of symplectomorphisms, which in the case considered is $SL(2,\mathbb{Z})$. In this paper we analyze the gauged supermembrane arising from the gauging of the abelian subgroups of this $SL(2,\mathbb{Z})$ group which has an intrinsic meaning in the theory. The gauging is automatically achieved by formulating the supermembrane with central charges as sections of a symplectic torus bundle with monodromy. The monodromy is also intrinsically defined by considering representations of $\Pi_1(\Sigma)$, the fundamental group of the Riemann base manifold of genus one ($\Sigma$), onto $\Pi_0(G)$ the isotopy group of the symplectomorphisms group $G$. The abelian subgroup of $SL(2,\mathbb{Z})$ acts naturally on the homology of the target torus (the fiber of the bundle \footnote{The complete fiber corresponds in this set-up to the target space, that in the case considered is $M_9\times T^2$ but the nontrivial topological properties are only associated to the compact sector.}) $H_1 (T^2)$. We identify, in our formulation of the supermembrane, the elements of $H_1(T^2)$ with $(p,q)$ KK charges. Besides, the winding numbers are directly related to the cohomology of the base manifold $\Sigma$. For a given monodromy there is a one to one correspondence between the symplectic torus bundle with that monodromy and the elements of the coinvariant group of the monodromy \cite{khan}. These elements are equivalence clases of KK $(p,q)$ charges which we explicitly described for the elliptic, parabolic, and hyperbolic monodromies. We classified the symplectic torus bundles in terms of the coinvariant group of the monodromy. It turns out that at the level of the supermembrane what is relevant are the elements of the coinvariant group of a given monodromy group. The possible values of the (p,q) charges on a given symplectic torus bundle with that monodromy are restricted to the corresponding equivalence class defining the element of the coinvariant group associated to the bundle. We also analyse the presence of torsion elements in the cohomology of the base of the manifold or equivalently $Z_m\oplus Z_n$ groups as the coinvariant group of the monodromy. We also obtained, using the same geometrical setting, the gauging of the trombone symmetry. It is constructed from a nonlinear representation of $SL(2,\mathbb{Z})$ and gives rise to a different symplectic torus bundle in comparison to the previous constructions in terms of linear representations.

 We showed the existence of a new $Z_2$ symmetry that plays the role of T-duality in M-theory interchanging the winding and KK charges but leaving the hamiltonian invariant. We expect that all monodromies associated to type IIA will arise from the dual symplectic torus bundle obtained from this new T-duality symmetry. Consequently, we expect that the global geometrical formulation of supermembranes we are proposing will provide a unified origin of all type II gauged supergravities in 9D. We may then conjecture that the supermembrane becomes the M-theory origin of all type II nine dimensional supergravities.

 From this construction of the supermembrane on symplectic torus bundle one may identify directly corresponding gauged supergravities in 9D. Moreover, a given gauged supergravity can only interact with a corresponding supermembrane on a symplectic torus bundle associated to a coinvariant element of the same monodromy, otherwise, an inconsistency with the transition functions on the bundle will occur.
We also obtain the explicit gauge degree of freedom of the theory, discuss a gauge fixing procedure and obtain the residual symmetry once the monodromy has been assumed.

 Recently in type II String Phenomenology the role of M2-branes wrapping homological 2-cycles with torsion has been used as a M-theory realization of the so-called discrete gauge symmetries $Z_N$. These symmetries may have a potential number of bondages from the phenomenological viewpoint as for example to be discrete symmetries that can help to realize proton stability or help to suppress some dangerous operators. It has been conjectured that this M2-branes at low energies would produce Bohm-Aranov particles \cite{angel}-\cite{luis}. In our constructions many of the M2-branes fiber bundles naturally are wrapped on homological 2-cycles with torsion. It would be interesting to see whether in compactifications down to 4D, it could be a  possible connection with our construction.

\section{Acknowledgements} We want to thank E. Bergshoeff, Y.
Lozano, I. Martin,  P. Meessen,  T.Ortin and  A. Vi\~na for helpful comments or discussions at different stages of this project. The work of MPGM is funded by the Spanish
Ministerio de Ciencia e Innovaci\'on (FPA2006-09199) and the
Consolider-Ingenio 2010 Programme CPAN (CSD2007-00042). The work of A. R. is funded by Proyecto FONDECYT 1121103.

\section{Appendix A}
 We are going to determine $\Lambda_0$. Without loss of generality we may assume $l_1$ and $m_1$ to be relatively prime integers. We have $\det(\mathbb{W})=n$
 It is important to notice that $\begin{pmatrix} p_1\\ q_1 \end{pmatrix}$ are also relatively prime integers. There always exists $\Lambda_0\in SL(2,\mathbb{Z})$ such that
 \bea
 \begin{pmatrix} l_1\\ m_1 \end{pmatrix}= \Lambda_0 \begin{pmatrix} p_1\\ q_1 \end{pmatrix}
 \eea

 We thus have from (\ref{t1}): \bea
 \begin{pmatrix} \widetilde{p}_1\\ \widetilde{q}_1 \end{pmatrix}= \begin{pmatrix} l_1\\ m_1 \end{pmatrix}.
 \eea
 We now introduce
 \bea
 \begin{pmatrix} r_2\\ r_1 \end{pmatrix}= \Lambda^{-1}\begin{pmatrix} l_2\\ m_2 \end{pmatrix}.
 \eea

Now we define $\mathbb{A}=\begin{pmatrix} p_1 & r_2\\q_1 & r_1 \end{pmatrix}$ consequently $
\mathbb{A} = \Lambda^{-1}\begin{pmatrix}l_1 & l_2\\ m_1 & m_2 \end{pmatrix}$, with $det \mathbb{A}=n$.
 We notice that $
 det\begin{pmatrix}\widetilde{l}_1 & \widetilde{l}_2 \\ \widetilde{m}_1 & \widetilde{m}_2 \end{pmatrix}=\mathbb{A}.
 $ We thus have a transformation interchanging winding and KK modes. The expression for $\Lambda_0$ may be obtained in the following way: There always exists integers $(b_2, b_1,d_1, c_1)$ such that there are $\mathbb{B}=\begin{pmatrix} p_1 & b_2 \\ q_1 & b_1 \end{pmatrix}$, and $\mathbb{C}=\begin{pmatrix} l_1 & d_1 \\ m_1 & c_1 \end{pmatrix}$, with
 \bea \begin{pmatrix} p_1 \\ q_1 \end{pmatrix}=\mathbb{B}\begin{pmatrix} 1 \\ 0 \end{pmatrix}, \quad
 \begin{pmatrix} l_1\\ m_1 \end{pmatrix}=\mathbb{C}\begin{pmatrix} 1 \\ 0 \end{pmatrix},
  \eea
 where $\mathbb{B}, \mathbb{C}\in SL(2,\mathbb{Z})$. Finally we can determine the transformation matrix $\Lambda_0$ . It corresponds to,
  \bea
 \Lambda_0= \begin{pmatrix} l_1 & d_1\\ m_1 & c_1 \end{pmatrix}\begin{pmatrix} p_1 & b_2 \\ q_1 & b_1 \end{pmatrix}^{-1}.
 \eea and together with the (\ref{za}) condition implies that the T-dual transformation (\ref {tt1}) $(\textrm{T-duality})^2= \mathbb{I}$.
 
 \section{Appendix B} In this appendix we are going to prove that the matrix of (\ref{hip}) particularized to the values corresponding to $n=1$, $m>0$,  $\begin{pmatrix} -2m+1& 2m+1\\ -m & m+1 \end{pmatrix}= T^2S T^{m+1}STS$ is conjugate to $-T^{-3}S(T^{-2}S)^{m-5}$. We denote by $\sim$ two conjugate matrices. We then have
 \begin{equation}
 \begin{aligned} 
&T^2ST^{m+1}STS= -T^2ST^mS T^{-1}\sim -TST^{m}S\sim (-1)^m T(STS)^m\\ &
\sim T(T^{-1}ST^{-1})^m\sim (T^{-1}ST^{-1})^{m-1}T^{-1}S\sim (T^{-1}ST^{-1})^{m-2}S\\ 
& \sim (T^{-1}ST^{-1})^{m-4}T^{-1}\sim -T^{-1}ST^{-1}(T^{-1}ST^{-1})^{m-5}T^{-1}\\ 
& \sim -(T^{-1}ST^{-1})(T^{-1}ST^{-1})^{m-6}T^{-1}ST^{-2}\sim -T^{-3}S T^{-1}(T^{-1}ST^{-1})^{m-6}T^{-1}S\\
& = -T^{-3}S(T^{-2}S)^{m-5}.\qquad\qquad\Box \end{aligned}
 \end{equation}
 
 Where we have used $STS= -T^{-1}ST^{-1}$.

\end{document}